\documentclass[fleqn,usenatbib,useAMS]{mnras}

\usepackage{newtxtext,newtxmath}
\usepackage{enumerate}

\usepackage[T1]{fontenc}
\usepackage{ae,aecompl}

\usepackage{graphicx}	
\usepackage{amsmath}	
\usepackage{amssymb}	
\usepackage{natbib}
\usepackage{epsfig}

\title[A Diffuse \lbrack NII\rbrack \,Halo in the Centaurs Cluster]{Discovery of a diffuse optical line emitting halo in the core of the Centaurus cluster of galaxies: Line emission outside the protection of the filaments.}

\author[S. L. Hamer et al.]
{\parbox[h]{\textwidth}
{
S. L. Hamer,$^{1}$\thanks{E-mail: stephen.hamer.astro@gmail.com}
A. C. Fabian,$^{1}$
H. R. Russell,$^{1}$
P. Salom\'e,$^{2}$
F. Combes,$^{2,3}$
V. Olivares,$^{2}$
F. L. Polles,$^{2}$
A. C. Edge,$^{4}$
R. S. Beckmann$^{5}$}
\\
\vspace*{6pt}\\
$^{1}$Institute of Astronomy, University of Cambridge, Madingley Road, Cambridge CB1 0HA, UK\\
$^{2}$LERMA, Observatoire de Paris, CNRS, PSL University, Sorbonne University, 75014 Paris, France\\
$^{3}$College de France, 11 Place Marcelin Berthelot, 75005 Paris, France\\
$^{4}$Institute for Computational Cosmology, Department of Physics, Durham University, South Road, Durham DH1 3LE, UK\\
$^{5}$Institut d'Astrophysique de Paris, CNRS \& Sorbonne Universit{\'e}, UMR 7095, 98bis Boulevard Arago, 75014, Paris, France\\
}

\date{Accepted XXX. Received YYY; in original form ZZZ}

\pubyear{2015}

\begin{document}
\label{firstpage}
\pagerange{\pageref{firstpage}--\pageref{lastpage}}
\maketitle

\begin{abstract}

\noindent We present the discovery of diffuse optical line emission in the Centaurus cluster seen with the MUSE IFU. The unparalleled sensitivity of MUSE allows us to detect the faint emission from these structures which extend well beyond the bounds of the previously known filaments.  Diffuse structures (emission surrounding the filaments, a northern shell and an extended Halo) are detected in many lines typical of the nebulae in cluster cores ([NII]$_{\lambda 6548\&6583}$ ,[SII]$_{\lambda 6716\&6731}$, [OI]$_{\lambda 6300}$, [OIII]$_{\lambda 4959\&5007}$ etc.) but are more than an order of magnitude fainter than the filaments, with the faint halo only detected through the brightest line in the spectrum ([NII]$_{\lambda 6583}$). These structures are shown to be kinematically distinct from the stars in the central galaxy and have different physical and excitation states to the filaments. Possible origins are discussed for each structure in turn and we conclude that shocks and/or pressure imbalances are resulting in gas dispersed throughout the cluster core, formed from either disrupted filaments or direct cooling, which is not confined to the bright filaments.

\end{abstract}




\section{Introduction}

X-ray observations of the cores of many massive galaxy clusters show highly peaked surface brightness profiles suggesting that the hot gas they contain is rapidly losing energy and should thus cool quickly to produce cool gas reservoirs \citep[see reviews by][]{fab12,mn12}.  
At the centre of these cool core clusters resides a massive brightest cluster galaxy (BCG)
which is often surrounded by an extended optical line emitting nebula \citep[e.g.][]{hec89,con01,ham16} 
in the form of filaments extending out over several kpc  \citep[e.g.][]{mcd10,mcd11}.
High resolution X-ray observations of cool core clusters show similar filamentary structures are also present within the hot gas of the intracluster medium (ICM). Comparison of these structures with H$\alpha$ filaments show there is typically a strong qualitative consistency, with line emitting structures matched by similar structures in the soft X-ray gas \citep[for example][]{fab03a,fab08}.  However, the reverse is not always true and ICM structures seen in X-rays do not always show line emitting counter parts.

Substantial work has been conducted studying the optical nebula within the centres of cool core clusters over the last few decades \citep[e.g.][etc.]{hec89,cra99,cra05,con01,ode04,ode10,jaf05,oon10,mcd11} revealing a wide range of extents and surface brightnesses between the different nebulae, though in most cases high surface filling factors are found within the central regions ($\sim$4\,kpc from the BCG nucleus) of the nebulae.  
NGC 4696, the BCG at the centre of a local (D$_L$\,=49.3\,Mpc) cool core galaxy cluster (the Centaurus cluster), is surrounded by a bright (L$_{H\alpha}$=1.6$\times$10$^{40}$\,erg\,s$^{-1}$) and well studied optical line emitting nebula \citep[e.g.][]{fab82,fab16,can11} making it an ideal target for deep studies of these objects.
The nebula is seen to take the form of a network of optical line emitting filaments extended over $\sim$14\,kpc within the cluster core to the south, west and east of the BCG \citep{can11,fab16}. However, emission has not been detected to the north of the BCG beyond $\sim$2\,kpc making this an unusual case.  

The optical filaments surrounding NGC 4696 correlate with similar structures seen in the soft X-ray emission \citep{cra05}. These filamentary structures occupy a region of the ICM with low temperatures and relatively low pressures as seen in \citet{snd16}. Cavities are also seen in the X-ray image which are inflated by radio jets from the AGN at the centre of the BCG \citep{fab05}. The energy injected into the ICM through this process is believed to be responsible for replacing much of the energy lost through X-ray emission \citep[see][for reviews]{fab12,mn12}.  This suggests the filaments may have formed as gas was dragged out of the BCG by the expanding radio jets. However, \citet{snd16} detect cavities to the north of the BCG in a region of high temperature and pressure where no filaments are seen.  This wide array of features in such a local southern hemisphere cluster make Centaurus an ideal system in which to study the cores of galaxy clusters.

One striking feature of the core of the Centaurus cluster when observed in soft X-ray and optical bands is its distorted morphology.  The H$\alpha$ filaments appear to extend only from the southern edge of the BCG \citep[see HST narrowband image in][for example]{fab16} and spiral outwards in a clockwise direction.  The X-ray morphology is similar extending the spiral structure further to the north east. There is an offset between the nucleus of the BCG and the X-ray centroid of the ICM seen in only a small fraction of clusters with optical nebulae \citep{san09,ham12} which suggests the BCG may not be at rest with respect to the cluster core. The ICM also shows evidence of cold fronts seen as an an anticorrelation between X-ray surface brightness and temperature \citep{snd16}. All of this suggests that the cluster core is sloshing which may be responsible for producing the spiral like structure seen in the filaments. 

In this paper we present observations of the core of the Centaurus cluster with the Multi Unit Spectroscopic Explorer (MUSE) instrument on the VLT which provides excellent sensitivity over the entire nebula.  In $\S$\,2 the observations and data analysis are described. The detection of previously unknown diffuse components to the nebula is presented in $\S$\,3. Potential origins for these structures are discussed in   $\S$\,4 and conclusions are presented in $\S$\,5.
We assume a standard cosmology throughout with H$_0$=70\,km\,s$^{-1}$\,Mpc$^{-1}$ and $\Omega_m$=0.3, as such at the redshift of the Centaurus cluster (z=0.0114) 1" = 233\,pc.

\section{Observations and Data Reduction}

Observations of the Centaurus cluster were obtained using the Multi Unit Spectroscopic Explorer \citep[MUSE,][]{bac10} integral field unit (IFU) on board ESOs VLT during period 94 (Program 094.A-0859, PI: S. Hamer) on 8 December 2014 and 7 January 2015. The observations consisted of four pointings of 820s each, three were targeted at the Brightest Cluster Galaxy with a 5" dither and 90$^{\circ}$ rotation between each resulting in a total field of view of $\sim$70$\times$70\,arcsec.  The fourth was targeted at an empty region $\sim$2' to the north west to provide an accurate measure of the sky spectrum which is uncontaminated by emission lines from the cluster.  The mean DIMM seeing during the observations was 1.2\,arcsec. 

The data were reduced using version 1.0.4 of the MUSE data reduction pipeline with the individual recipes being executed from the European Southern Observatory Recipe Execution Tool (ESOREX v. 3.12) command-line interface.  Sky subtraction of the final datacube was done using version 2.1 of the Zurich Atmosphere Purge \citep[ZAP,][]{sot16} measuring the sky from the offset region before subtracting it from the datacube.  The data are corrected for galactic extinction using an O'Donnell extinction curve \citep{odl94} and the dust maps of \citet{sfd98} \citep[which were recalibrated in][]{sfd11}. Spectra were fitted using the latest version of {\sc ppxf} \citep{cap17}, taking stellar templates from the Indo-US spectral library \citep[with T$_{eff}\sim$\,3000\,--\,30000\,K and \lbrack Fe/H\rbrack$\sim$\,-3\,--\,1,][]{val04} and fitting continuum and nebular emission simultaneously.  Local extinction was accounted for using ppxfs built in function to include a reddening component to the spectral fit.

In regions of weak line emission but strong continuum the Balmer absorption features from stars completely dominate over the gas emission lines making it extremely difficult to accurately fit the Balmer features.  It should be noted that while there is no reason to expect a bright stellar component to be spatially and kinematically correlated with the gas emission the full region covered by the observation falls within the BCGs extended stellar envelope.  This results in the Balmer emission lines falling close to the bottom of the intrinsically broad Balmer absorption features in the stellar spectrum except in cases where line emission has a significant velocity offset from the BCG. At times this resulted in a degeneracy between the absorption and emission, in which overly deep absorption features were fit, balanced by extremely broad emission features at the same velocity.

In order to eliminate this problem the fitting was performed in a three step process.  First the wavelengths within $\pm$300\,km\,s$^{-1}$ of the expected position of each bright emission line were masked and a continuum only fit performed to the remaining parts of the spectra.  After this the resulting continuum models were subtracted from the full spectra and a fit to the residual emission line features was performed.  Finally a combined fit (continuum and emission lines) to the full spectrum was performed fixing the kinematics of the emission lines to those found from the emission line only fit. This process produced excellent fits to the spectrum, while maintaining the simultaneous fitting of continuum and emission features and minimising the degeneracy between absorption and broad emission lines. 



\section{Results}
\label{sec:res}
\begin{figure*}
\psfig{figure=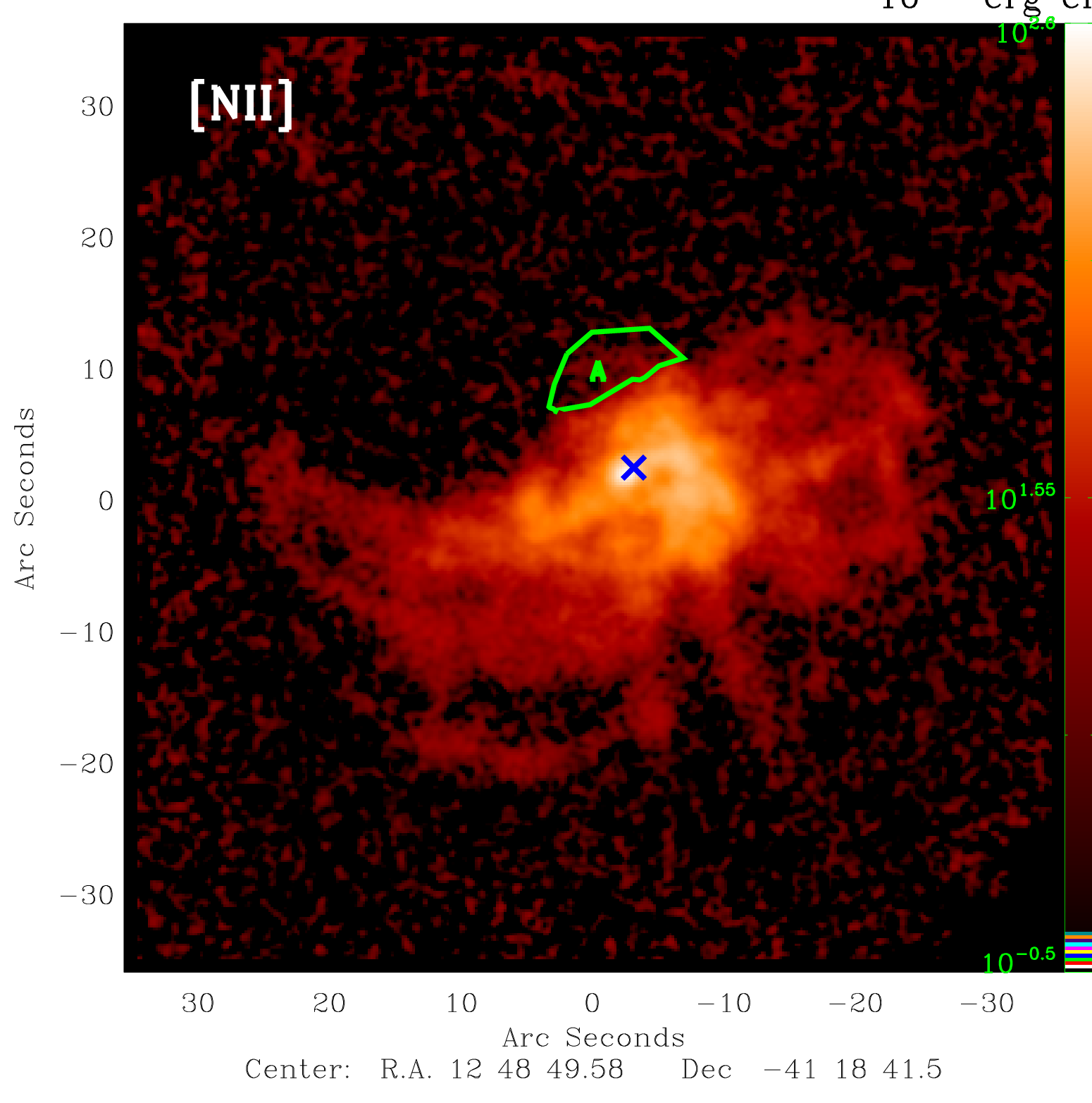,width=8.5cm}
\psfig{figure=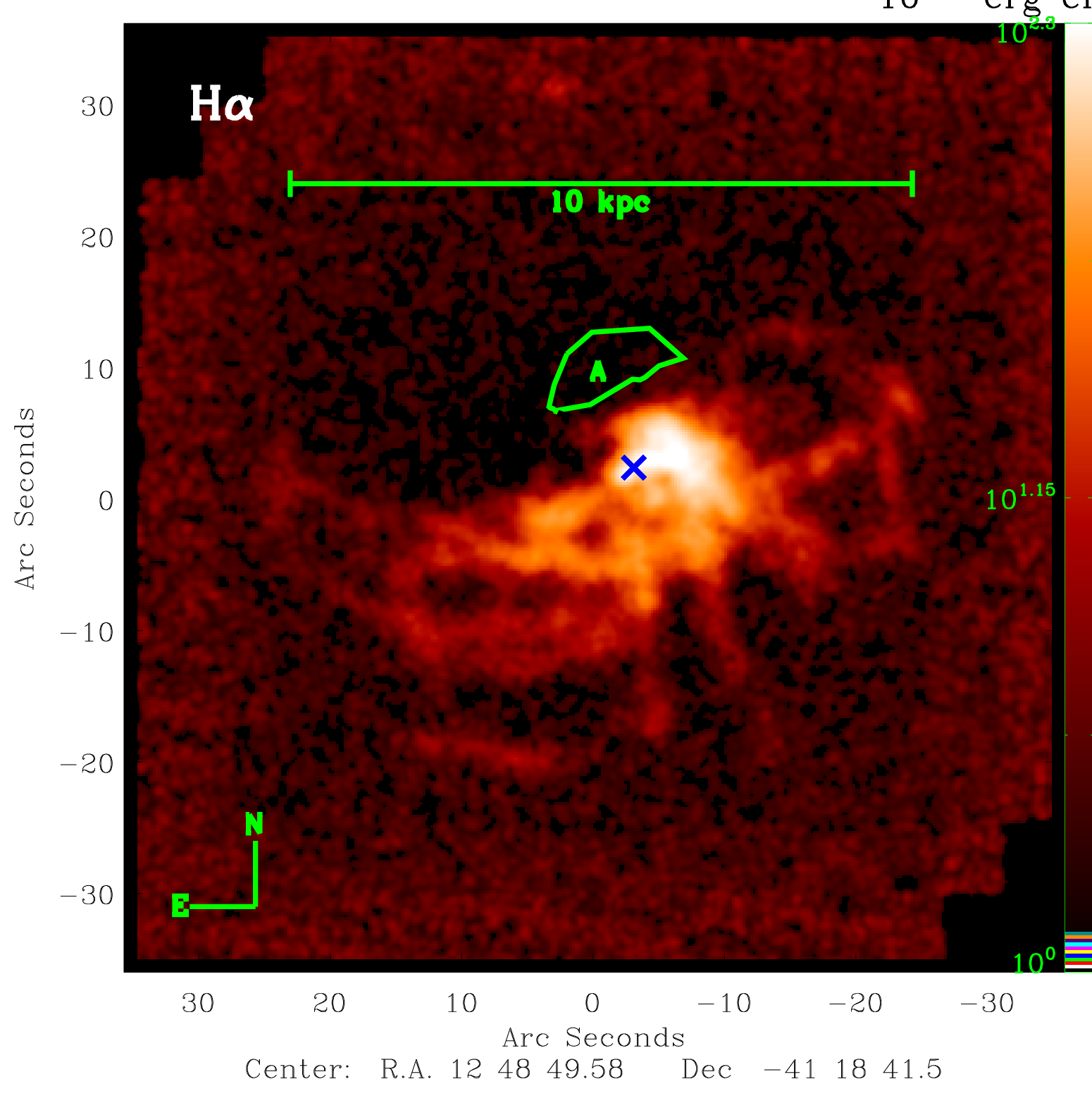,width=8.5cm}
\vspace{0.5cm}
\caption[H$\alpha$ and \lbrack NII\rbrack$_{\lambda 6583}$ narrow band images]{ Narrow band images produced from the MUSE data cube showing the [NII]$_{\lambda 6583}$  (left) and H$\alpha$ (right) emission.  [NII]$_{\lambda 6583}$ is the brighter line and clearly shows a more extended and diffuse structure than the purely filamentary H$\alpha$.  [NII]$_{\lambda 6583}$ emission is present between the filaments and to the north and north east of the cluster where no filaments are present. Region A is the same on each image and highlights such an area to the north east of the BCG.  The centre of the BCG is indicated by the position of the blue $\times$ in each image. The flux values are given per pixel with a pixel scale of 0.04 arcsec$^2$ (0.2"$\times$0.2"). The 3$\sigma$ level is 10.8$\times$10$^{-20}$ and 16.1$\times$10$^{-20}$ erg\,cm$^{-2}$\,s$^{-1}$\,\AA$^{-1}$ for the [NII]$_{\lambda 6583}$ and H$\alpha$ image respectively while the total [NII]$_{\lambda 6583}$ flux density from region A is 1444.2$\times$10$^{-20}$ erg\,cm$^{-2}$\,s$^{-1}$\,\AA$^{-1}$.
}
\label{fig:nbs}
\end{figure*}

Continuum subtracted narrow band images produced from the datacube detect the previously known filamentary H$\alpha$ emitting nebula (Figure \ref{fig:nbs} {\em Right}). This filamentary nebular has been extensively studied over the last three decades using both narrow band imaging \citep[e.g.][]{spa89,cra05,fab16} and IFU observations \citep[e.g. the long VIMOS observation of][]{can11}.  However, while the majority of the [NII]$_{\lambda 6583}$ emission also traces this filamentary structure, we note that it appears more diffuse, with [NII]$_{\lambda 6583}$ emission extending between and beyond the filaments (Figure \ref{fig:nbs} {\em Left}).  As both of these lines typically fall within the bands used by imaging studies this contrast is not apparent in previous observations of this cluster. 

\begin{figure*}
\psfig{figure=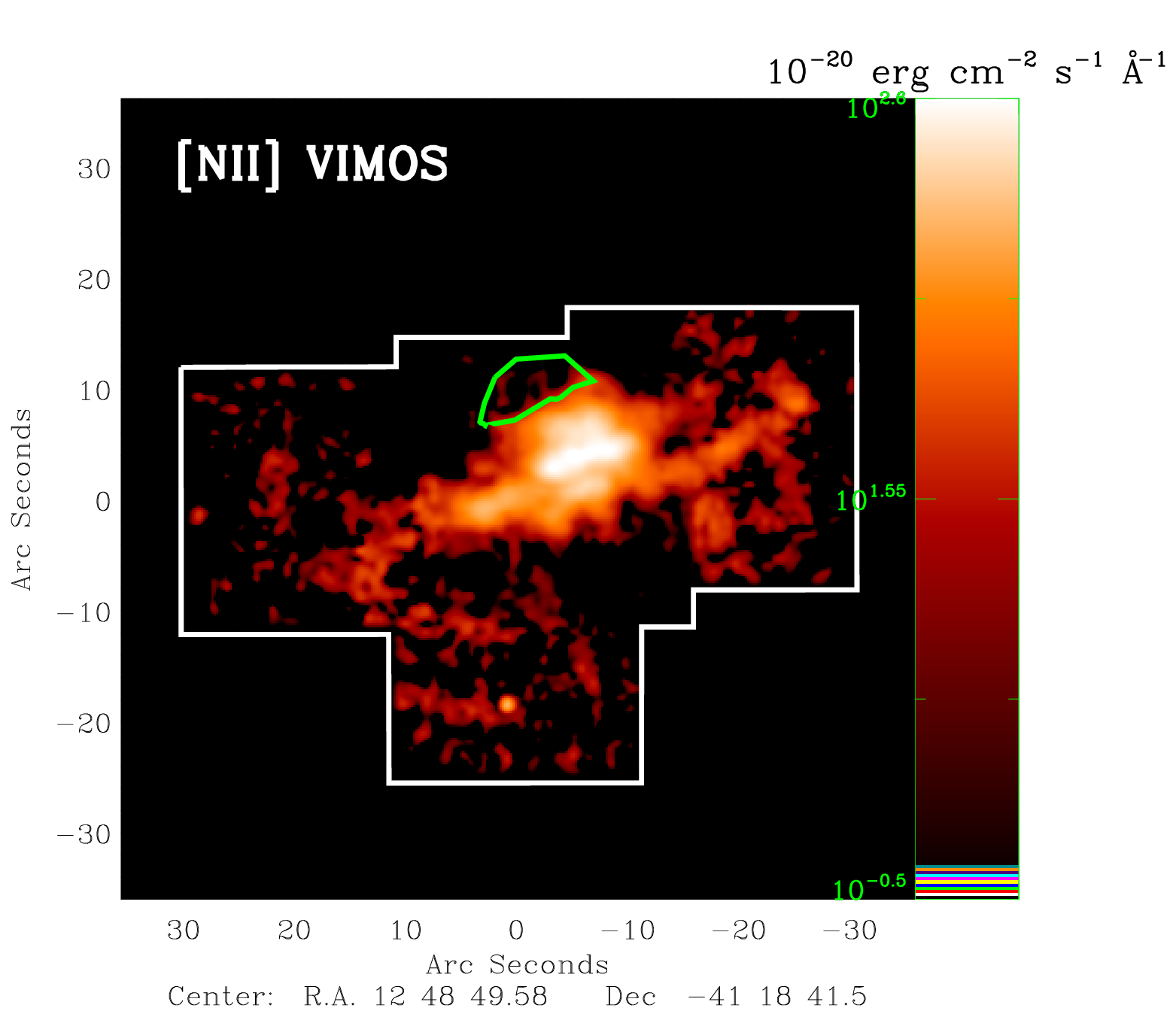,width=9cm}
\psfig{figure=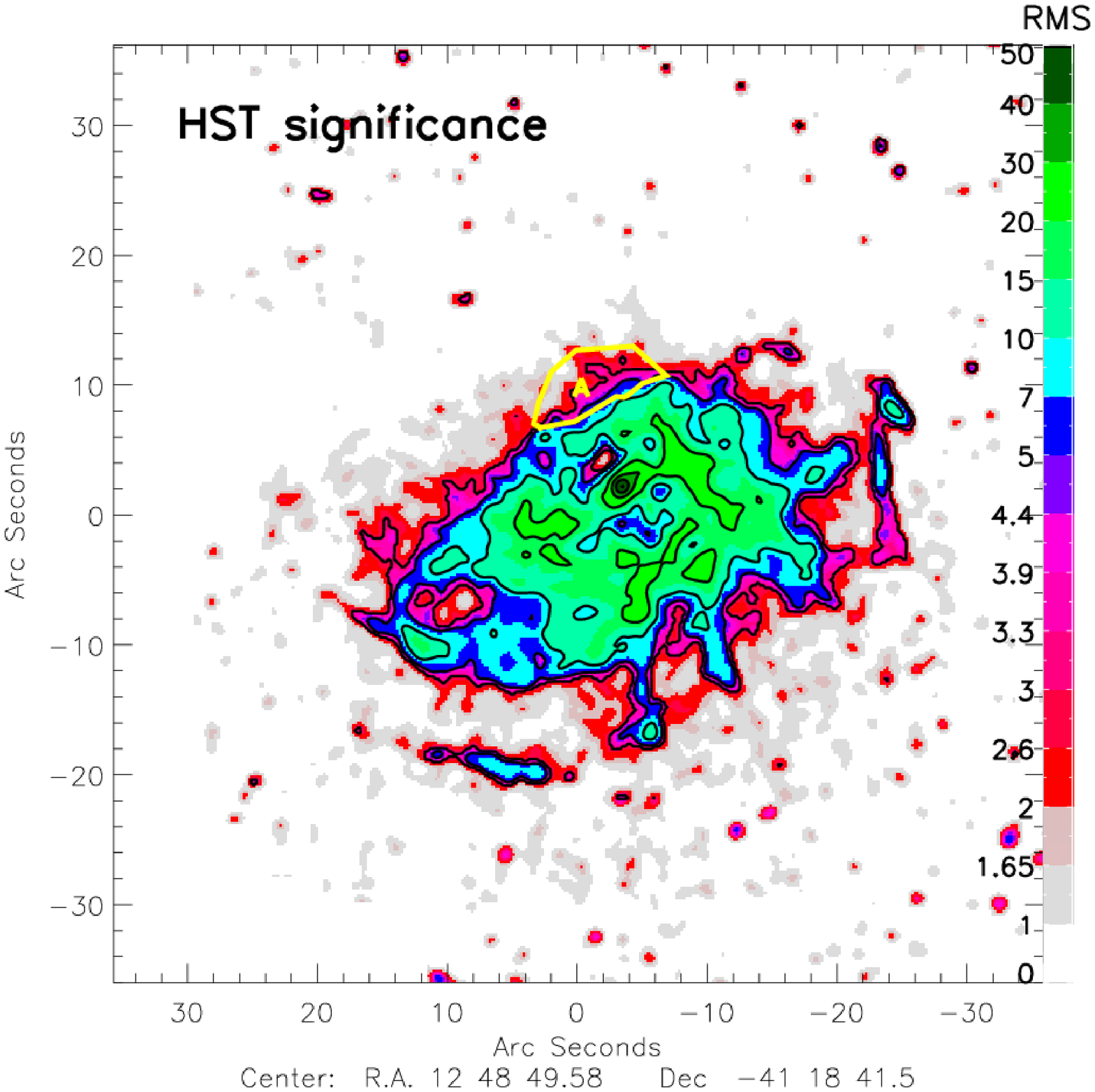,width=8.5cm}
\vspace{0.5cm}
\caption[Diffuse component comparison to previous data]{ {\em Left:} Narrow band [NII] image produced from the VIMOS data cube (resampled to match the MUSE pixel scale of 0.04 arcsec$^2$ or 0.2"$\times$0.2") of \citet{can11} showing the [NII]$_{\lambda 6583}$ emission from the same wavelength range as shown for MUSE in figure \ref{fig:nbs}.  Flux values are given per pixel. The limits of the combined field of view for these observations is show in white and region A is outlined in green.  Emission can be seen from this region, although at very low significance compared to the MUSE data.  The 3$\sigma$ level is 52.5$\times$10$^{-20}$ erg\,cm$^{-2}$\,s$^{-1}$\,\AA$^{-1}$ and the total flux from region A is 26.1$\times$10$^{-20}$ erg\,cm$^{-2}$\,s$^{-1}$\,\AA$^{-1}$ {\em Right:}  Significance map of the $HST$ narrow band image from \citet{fab16} measured in units of the image RMS.  Region A is shown for comparison.  While there is evidence of emission from this region most of it has a significance of less than 3$\sigma$.  
}
\label{fig:hst_sig}
\end{figure*}

Figure \ref{fig:hst_sig} (Left) shows a narrowband image produced from the VIMOS \citep[VIsible Multi-Object Spectrograph,][]{lef03} data of \citet{can11} using the same wavelength limits as the MUSE [NII] narrow band image seen in figure \ref{fig:nbs}.  The improvement in sensitivity from the MUSE data is immediately apparent when comparing these two images.  Not only is the diffuse component from the nebula not apparent in the VIMOS  data, but some of the filaments are not visible in this narrowband image either (e.g. the centre filament extending to the south-west of the centre). What is also apparent from comparing these two images is the diffuse structure to the north of the BCG (region A) seen in the MUSE [NII] narrow band image is not clearly detected in the VIMOS narrow band image.  In the MUSE [NII] narrow band image region A has an average surface brightness of 51.9$\times$10$^{-20}$ erg\,cm$^{-2}$\,s$^{-1}$\,\AA$^{-1}$\,arcsec$^{-2}$ far in excess of the background level of 0.17$\times$10$^{-20}$ erg\,cm$^{-2}$\,s$^{-1}$\,\AA$^{-1}$\,arcsec$^{-2}$, however in the VIMOS narrow band image the surface brightness of region A (64.7$\times$10$^{-20}$ erg\,cm$^{-2}$\,s$^{-1}$\,\AA$^{-1}$\,arcsec$^{-2}$) is less than twice that of the background (36.9$\times$10$^{-20}$ erg\,cm$^{-2}$\,s$^{-1}$\,\AA$^{-1}$\,arcsec$^{-2}$) indicating it is not sufficiently bright to stand out above the noise.

The $HST$ narrow band image of \citet{fab16} is one of the deepest to be taken on this object.  It shows some tentative evidence of the line emission seen in region A however, a direct comparison of this region to a significance map of the $HST$ image (Figure \ref{fig:hst_sig}, Left) suggests that much of this does not exceed a 3$\sigma$ significance.  Comparing the [NII] and H$\alpha$ narrow band images from MUSE it is apparent that while region A shows the presence of [NII] emission no evidence of H$\alpha$ emission is seen (the H$\alpha$ surface brightness is 0.25$\times$10$^{-20}$ erg\,cm$^{-2}$\,s$^{-1}$\,\AA$^{-1}$\,arcsec$^{-2}$ barely exceeding the background level of 0.24$\times$10$^{-20}$ erg\,cm$^{-2}$\,s$^{-1}$\,\AA$^{-1}$\,arcsec$^{-2}$).  While it is possible that this region is really lacking in H$\alpha$ emission this would suggest a different excitation mechanism or significantly different metallicity than the rest of the nebula.  The more plausible explanation is that H$\alpha$ emission is present in this region but is lost to the Balmer absorption feature of the BCGs stellar continuum. This structure is not discussed in previous studies of this cluster and most suggest that the nebular extends exclusively on the southern side of the BCG.

This faint diffuse emission is detected thanks to the combination of the unprecedented sensitivity of MUSE and its ability to isolate individual emission lines \citep[e.g.][]{ham15}.  This is particularly important in this case where the underlying continuum spectrum has significant structure in the form of the stellar H$\alpha$ absorption feature.  By isolating a single line when producing the narrow band image the continuum variations within the band are minimised and the contribution of the continuum emission is significantly reduced.  This reduces the errors associated with the continuum subtraction which allows for the faint structures to be seen at much higher contrast than is possible with available narrowband images \citep[e.g.][]{fab16}.

\begin{figure}
\psfig{figure=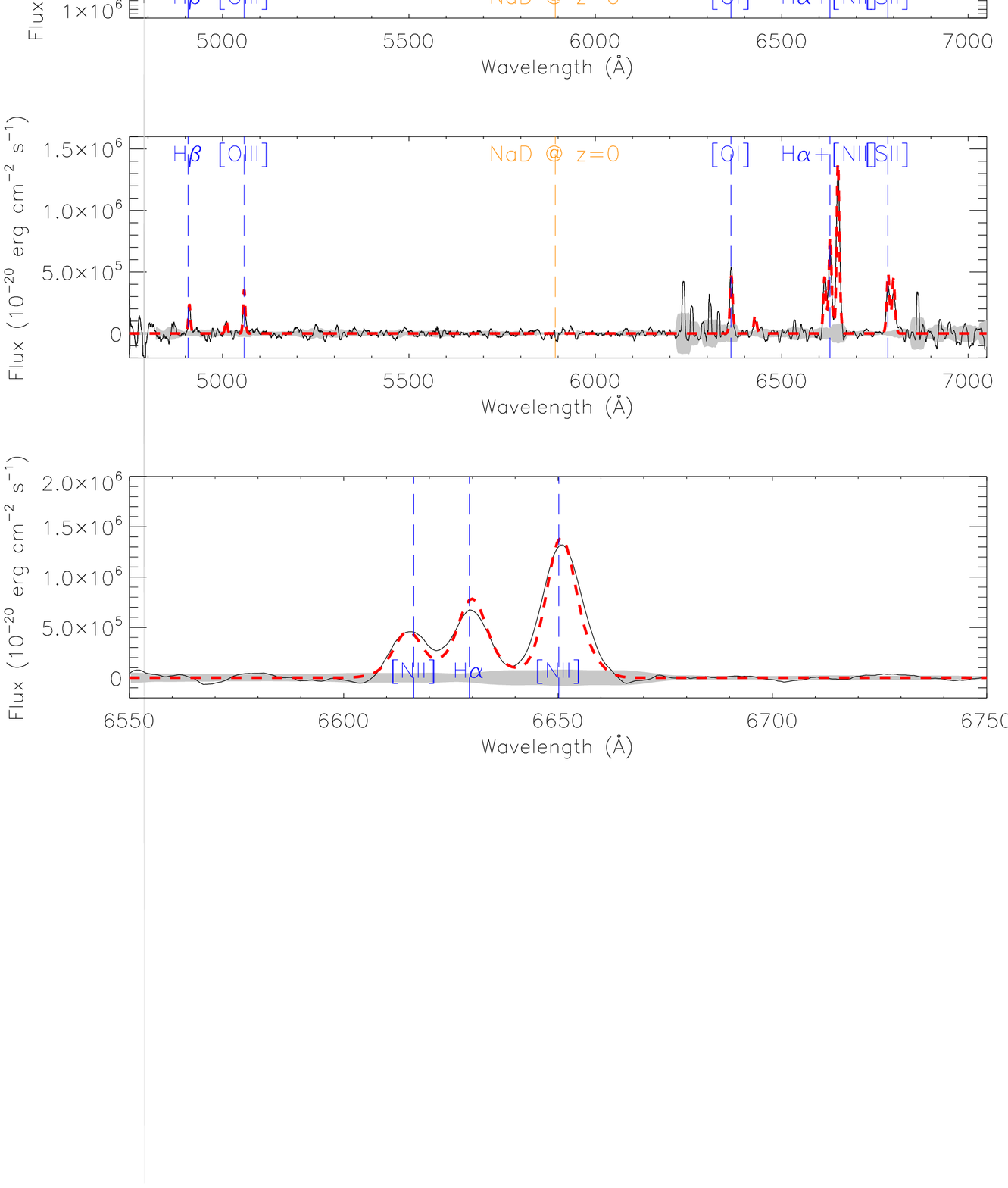,width=8.5cm}
\vspace{-0.75cm}
\caption[Spectrum and fit from the total nebula]{{\em Top:} Integrated spectrum extracted from the full cube masking only bad spaxels in which no data was present.  {\em Centre:} Continuum subtracted integrated spectrum.  {\em Bottom:} Continuum subtracted integrated spectrum, showing only the region around [NII] and H$\alpha$.  The red dashed lines show the fits to the continuum (top) and emission line (centre and bottom) features of the spectrum. The grey shaded region shows the 1$\sigma$ noise on the spectrum as a function of wavelength. Note the detection of multiple optical emission lines in this spectrum.}
\label{fig:tot_spec}
\end{figure}

The top panel of figure \ref{fig:tot_spec} shows the total integrated spectrum produced from the datacube along with the fit to the stellar continuum.  Most strong optical emission lines within the covered wavelength are already visible without continuum subtraction.  The middle and bottom panels of figure \ref{fig:tot_spec} show the continuum subtracted spectra (full and zoomed in around the H$\alpha$-[NII] complex) along with the fits to the strong emission lines present.
[NII]$_{\lambda 6583}$ is by far the strongest line in the spectrum with a total detected flux of 27.2$\pm$0.57$\times$10$^{-14}$\,erg\,cm$^{-2}$\,s$^{-1}$. This is a factor of 2.33$\pm$0.06 brighter than the H$\alpha$ flux of 11.7$\pm$0.17$\times$10$^{-14}$\,erg\,cm$^{-2}$\,s$^{-1}$, which is also more greatly affected by stellar absorption.  These two factors make the H$\alpha$ narrow band image in figure \ref{fig:nbs} less reliable than the [NII]$_{\lambda 6583}$ narrow band image for determining the overall structure of the nebula. 

As such the morphology seen in the [NII]$_{\lambda 6583}$ emission, with the additional more extended and seemingly diffuse structures (see Figure \ref{fig:nbs}), likely represents the true extent and structure of the nebula.
These diffuse structures are too faint to have been studied with previous observations so the MUSE data enables a quantitative study of their properties for the first time.
  Much of the extended emission follows the approximate structure of the bright filaments (with the exception of region A) suggesting the structures may be related.  However, this extended emission is more than an order of magnitude fainter than the faintest filaments in this system and extends far beyond the tightly constrained filaments seen in high resolution HST imaging \citep{fab16}.  This suggests the possibility they are a distinct, previously unknown, component of the nebula which, while seemingly related to the filaments, may have different formation and excitation mechanisms.

\subsection{The filaments}

The filamentary system in Centaurus has been extensively studied by many authors \citep[e.g.][to name a few]{spa89,cra05,fab16,snd16}. In particular \citet{can11} perform a detailed analysis using deep VIMOS IFU observations to study the kinematics and physical properties of the optical line emitting filaments.  The MUSE data presented in this paper are considerably more sensitive than this VIMOS data and have improved spatial sampling and a much larger field of view (see Figure \ref{fig:hst_sig}). However, the filaments in this system are bright and were already well detected by the VIMOS observations. The kinematic and physical properties measured with MUSE are consistent with those found in \citet{can11} and thus will not be readdressed in this paper.  

We define the filaments as the structures detected at greater than 3$\sigma_{rms}$ in the H$\alpha$ narrowband image (Figure \ref{fig:nbs}). Total detected fluxes of all measured lines from this filament region are given in Table \ref{tab:fil}. The brightest lines in the spectrum ([NII]$_{\lambda6548\&6583}$ \& H$\alpha$) have fluxes slightly higher than those quoted by \citet{can11} but this can be attributed to recovering the filaments over a larger extent than was possible with the VIMOS data.  Likewise the H$\beta$ and [OIII]$_{\lambda4959\&5007}$ fluxes are significantly higher as these are now recovered from the full filament system not just the central 27$\times$27\,arcsec$^2$.  The [OI]$_{\lambda6300}$ and [SII]$_{\lambda6716\&6731}$ are significantly brighter than in the VIMOS data, this is likely due to the increased sensitivity of MUSE allowing flux from these weaker lines to be recovered from the fainter regions of the filaments.

\begin{table}
\begin{center}
\centerline{\sc Table. \ref{tab:fil}.}
\centerline{\sc Total emission line fluxes from the filaments}
	\begin{tabular}{c c c}
    	\hline
		\noalign{\smallskip} 
        Emission & Total Flux & Flux/F$_{H\alpha}$ \\
        Line & (erg\,s$^{-1}$\,cm$^{-2}$) & \\
        \hline
        H$\beta_{\lambda4861}$ & 4.1$\pm$0.06\,$\times$\,10$^{-14}$ & 0.34 \\
        \lbrack OIII]$_{\lambda4959}$ & 1.1$\pm$0.33\,$\times$\,10$^{-14}$ & 0.09 \\
        \lbrack OIII]$_{\lambda5007}$ & 3.3$\pm$0.98\,$\times$\,10$^{-14}$ & 0.28 \\
        \lbrack NII]$_{\lambda5755}$ & $<$5.8\,$\times$\,10$^{-16}$ & 5\,$\times$\,10$^{-3}$ \\
        \lbrack OI]$_{\lambda6300}$ & 7.9$\pm$1.38\,$\times$\,10$^{-14}$ & 0.67 \\
        \lbrack NII]$_{\lambda6548}$ & 9.1$\pm$0.19\,$\times$\,10$^{-14}$ & 0.78 \\
        H$\alpha_{\lambda6562.8}$ & 11.7$\pm$0.17\,$\times$\,10$^{-14}$ & 1.00 \\
        \lbrack NII]$_{\lambda6583}$ & 27.2$\pm$0.57\,$\times$\,10$^{-14}$ & 2.33 \\
        \lbrack SII]$_{\lambda6716}$ & 9.9$\pm$0.12\,$\times$\,10$^{-14}$ & 0.85 \\
        \lbrack SII]$_{\lambda6731}$ & 8.6$\pm$0.11\,$\times$\,10$^{-14}$ & 0.73 \\
        \hline
        \end{tabular}
\caption[Total fluxes from the filaments]{Total fluxes of the measured emission lines from the filaments in the core of the Centaurus cluster corrected for galactic and local extinction.  The \lbrack NII\rbrack$_{\lambda5755}$ flux is a 3$\sigma$ upper limit as denoted by the preceding $<$.}
\label{tab:fil}
\end{center}
\end{table}

\begin{figure*}
\psfig{figure=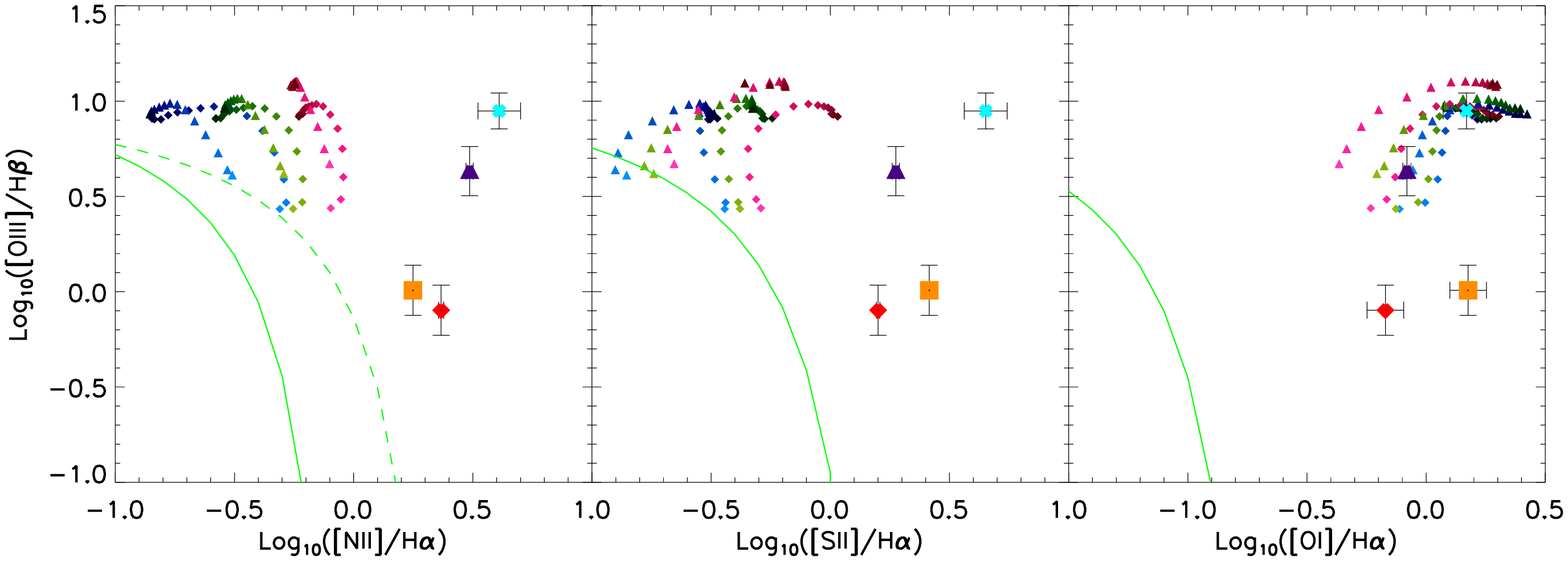,width=17.0cm}
\caption{Comparison of the line ratio excitation diagnostics for three distinct regions of the Centaurus optical nebula. The filaments are shown as the red diamond, the extended regions (between the filaments) as the purple triangle and the northern extent (region A) as the cyan cross.  All regions are inconsistent with star formation excitation models \citep[whose upper limit is shown by the green line][]{kew06}.  The particle heating (cosmic rays) model of \citet{fer09} is shown as the orange square. Shock models from \citep{all08} are shown for particle densities of n=100\,cm$^{-3}$ (small diamonds) and n=1000\,cm$^{-3}$ (small triangles). The colours represent shocks with differing magnetic field strength (blue=0.1, green=1.0, red=5.0 $\mu$G for n=100\,cm$^{-3}$ and blue=10, green=32, red=100 $\mu$G for n=1000\,cm$^{-3}$) while darker colours indicate faster shocks (in the range 300--1000\,km\,s$^{-1}$).  The filaments are most consistent with the particle heating but excitation by shocks slower than those considered cannot be ruled out in this case.  The line ratios of the extended emission and the northern extent are inconsistent with all models.}
\label{fig:bpt}
\end{figure*}

Figure \ref{fig:bpt} shows the standard line ratio diagnostic plots \citep{bpt81,kew06} for emission from the filaments (red diamond).  The position of the filaments on these falls within the LINER region of the plots, consistent with the findings of \citep{can11}. The line ratios are strongly inconsistent with models of star formation excitation \citep{kew06} as is typically  the case for optical nebulae in cluster cores \citep[e.g.][]{jf88,cf92,ham16}.  The line ratios are very consistent with the particle heating models (cosmic rays) produced by \citet{fer09} which are often invoked to explain the line ratios from optical nebula in galaxy cluster cores. However in this case we cannot rule out excitation through slow shocks \citep[$<$300\,km\,s$^{-1}$, the lower velocity limit for the shock models plotted in Figure \ref{fig:bpt}, taken from][]{all08} from this analysis alone.

\begin{figure}
\psfig{figure=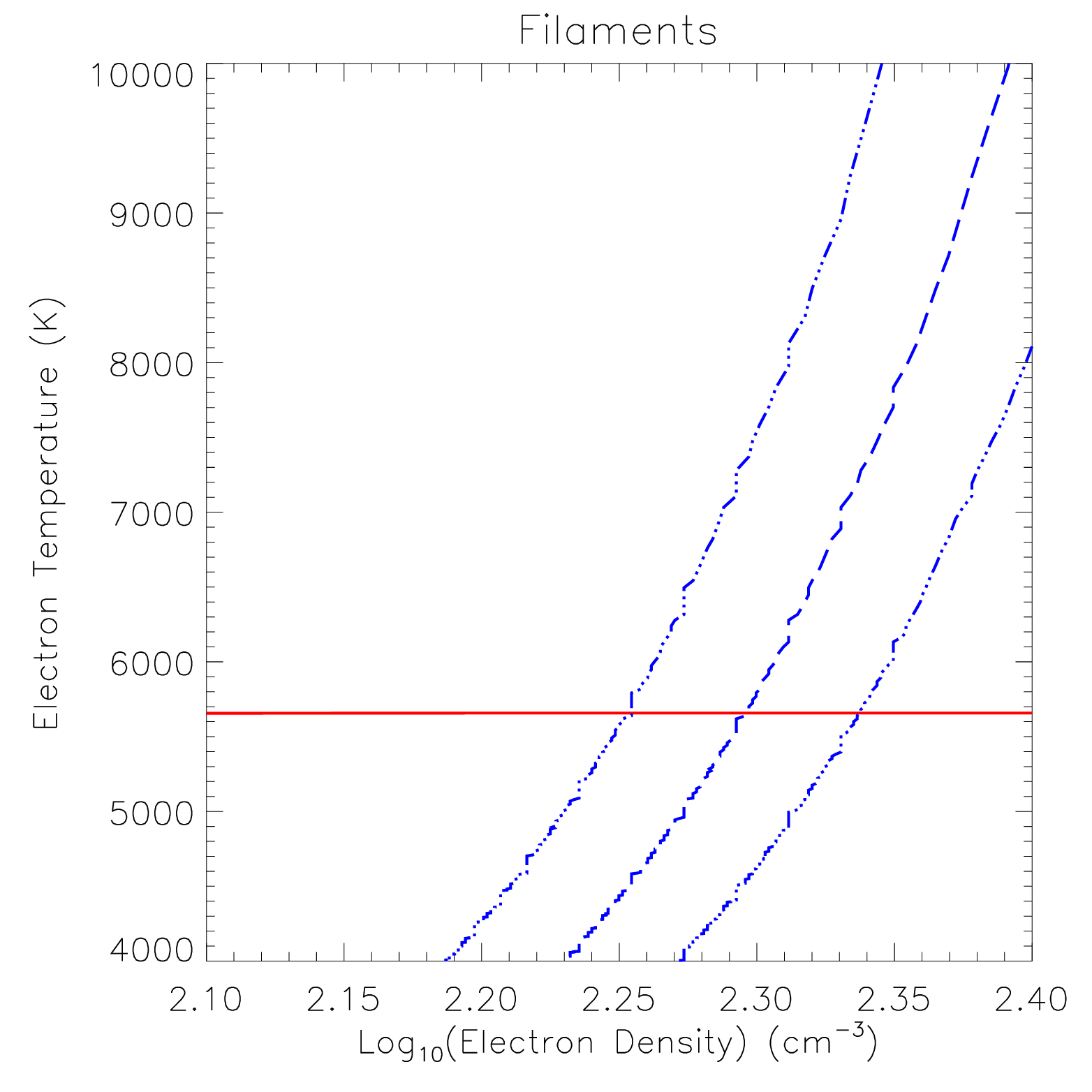,width=8.5cm}
\caption{Comparison plot of electron temperature and density within the filaments as measured from the [NII]$_{\lambda5755}$ / ([NII]$_{\lambda6548}$ + [NII]$_{\lambda6583}$) (red solid line) and [SII]$_{\lambda6716}$ / [SII]$_{\lambda6731}$ (blue dashed line) emission line ratios.  The electron density is well constrained by the [SII]$_{\lambda6716}$ / [SII]$_{\lambda6731}$ ratio with the 1$\sigma$ error bounds shown by the blue triple dot dashed lines. While the electron temperature cannot be measured as the [NII]$_{\lambda5755}$ line is undetected strong upper limits can be placed on the [NII]$_{\lambda5755}$ flux.  In this way the electron temperature can still be well constrained for this warm ionised gas.}
\label{fig:fil_temden}
\end{figure}

The [SII]$_{\lambda6716}$ / [SII]$_{\lambda6731}$ line ratio measured from the MUSE data (1.16$\pm$0.021) is slightly lower than that measured from the VIMOS data (1.22$\pm$0.13). The MUSE data also allow a considerably stronger upper limit to be placed on the flux of the [NII]$_{\lambda5755}$ emission line which is significantly weaker than the measured [NII]$_{\lambda6548}$ \& [NII]$_{\lambda6583}$ lines.  The combination of these five lines allows strong constraints to be placed on the electron temperature (T$_e$) and electron density (n$_e$) within the filaments. These diagnostics give an upper limit on electron temperature in the filaments of T$_e<$\,5685\,K with an electron density of n$_e$=159--198\,cm$^{-3}$ (Figure \ref{fig:fil_temden}) assuming a lower limit on temperature of 3500\,K.  This corresponds to an  upper limit on the pressure within the filaments of P$_e<$\,1.13$\times$10$^6$\,K\,cm$^{-3}$ which is consistent with the ICM pressure within the filaments over a similar extent \citep[0.06--0.07\,keV\,cm$^{-3}$ $\sim$ 0.7--0.8$\times$10$^6$\,K\,cm$^{-3}$, from Figure 17 in][]{snd16}.

\subsection{The diffuse extended nebula}

\begin{figure}
\psfig{figure=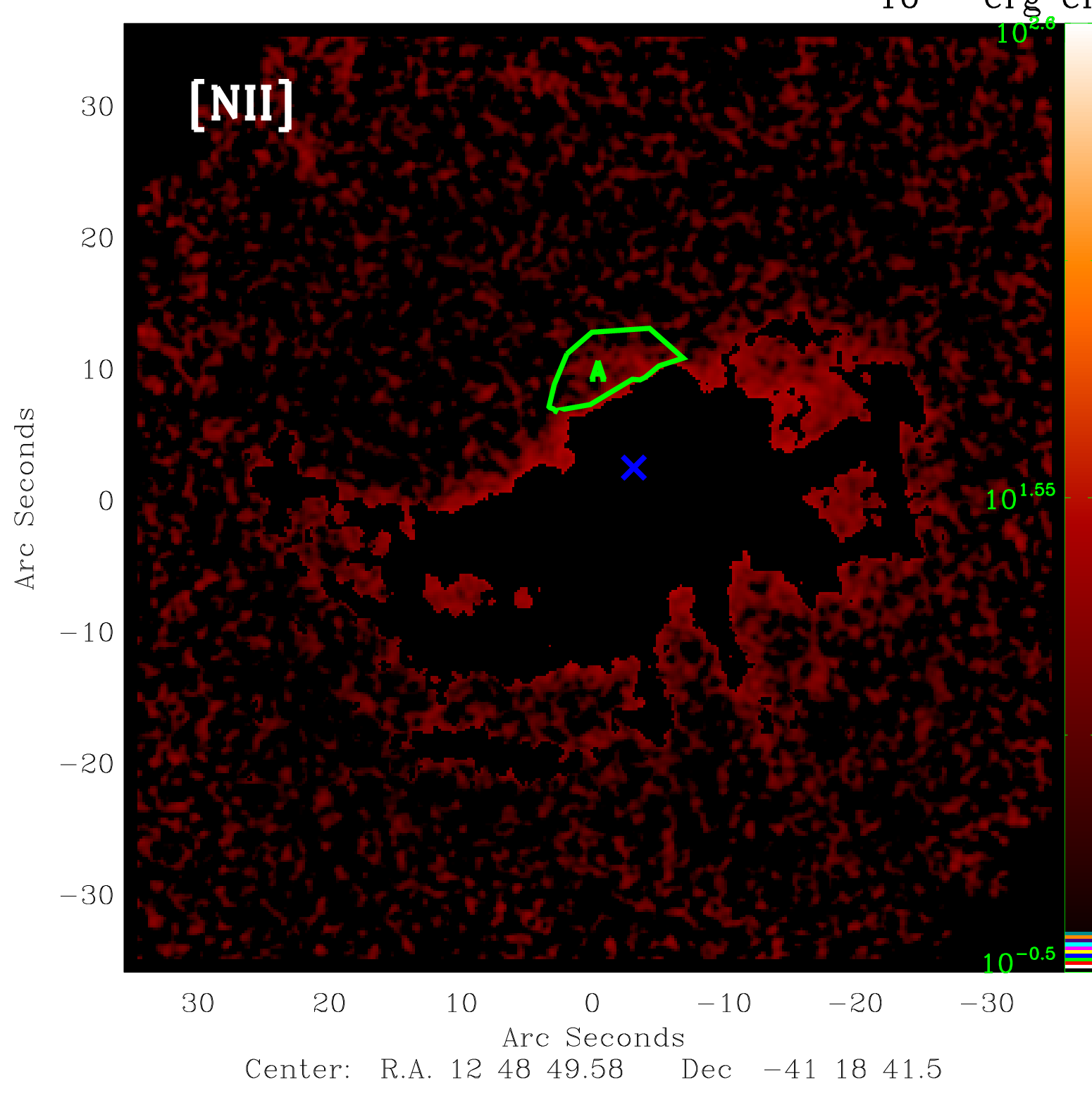,width=8.5cm}
\vspace{-0.75cm}
\caption[Masked \lbrack NII\rbrack image]{The \lbrack NII\rbrack$_{\lambda 6583}$ narrow band image from Figure \ref{fig:nbs} with regions containing H$\alpha$ flux at greater than 3$\sigma_{rms}$ masked to highlight the extended emission. The centre of the BCG is indicated by the blue cross and region A is labelled.}
\label{fig:nbs_msk}
\end{figure}

\begin{figure}
\psfig{figure=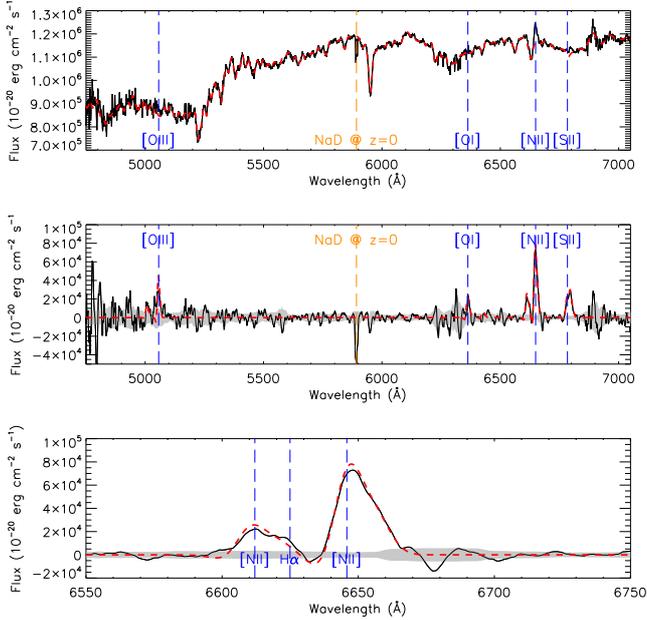,width=8.5cm}
\vspace{0.75cm}
\caption[Spectrum and fit from the extended nebula]{{\em Top:} Spectrum extracted from the full cube after masking spaxels which contain H$\alpha$ filaments as seen in the H$\alpha$ narrow band image. In effect the region of the datacube which does not contain H$\alpha$ filaments.  {\em Centre:} Continuum subtracted spectrum extracted from the region of the datacube which does not contain H$\alpha$ emission, in effect the non filamentary regions of the nebula. {\em Bottom:} Continuum subtracted spectrum of the non filamentary regions, showing only the spectral range around [NII] and H$\alpha$.  The red dashed lines show the fits to the continuum (top) and emission line (centre and bottom) features of the spectrum. The grey shaded region shows the 1$\sigma$ noise on the spectrum as a function of wavelength.  Note the strong [NII]$_{\lambda 6583}$ emission and additional lines present from this diffuse extended region despite the H$\alpha$ emission not being visible in the narrow band image.}
\label{fig:halo_spec}
\end{figure}

While evidence of the extended emission filling the regions between cavities is present in the data used by previous studies \citep[e.g.][]{can11,fab16} its properties have not been studied.  The resolution and sensitivity of the MUSE data allows these regions to be isolated and studied in detail so they can be contrasted to the brighter filaments.  We define this extended emission as coming from pixels in which [NII]$_{\lambda 6583}$ emission is detected at greater than 3$\sigma_{rms}$ but H$\alpha$ is not in the narrow band images (Figure \ref{fig:nbs}, a masked [NII]$_{\lambda 6583}$ image is provided in Figure \ref{fig:nbs_msk}).  The total spectrum from this emission, along with the spectral fits can be found in Figure \ref{fig:halo_spec}.  The [NII]$_{\lambda 6583}$ emission is immediately apparent in this spectrum, peaking at the red edge of the H$\alpha$ absorption feature even prior to continuum subtraction.  The total detected fluxes of all measured lines from this extended emission are given in Table \ref{tab:ext}.

\begin{table}
\begin{center}
\centerline{\sc Table. \ref{tab:ext}.}
\centerline{\sc Total emission line fluxes from the extended emission}
	\begin{tabular}{c c c}
    	\hline
		\noalign{\smallskip} 
        Emission & Total Flux & Flux/F$_{H\alpha}$ \\
        Line & (erg\,s$^{-1}$\,cm$^{-2}$) & \\
        \hline
        H$\beta_{\lambda4861}$ & 2.5$\pm$0.73\,$\times$\,10$^{-15}$ & 0.35 \\
        \lbrack OIII]$_{\lambda4959}$ & 3.6$\pm$0.20\,$\times$\,10$^{-15}$ & 0.50 \\
        \lbrack OIII]$_{\lambda5007}$ & 10.8$\pm$0.61\,$\times$\,10$^{-15}$ & 1.49 \\
        \lbrack NII]$_{\lambda5755}$ & $<$3.8\,$\times$\,10$^{-16}$ & 0.05 \\
        \lbrack OI]$_{\lambda6300}$ & 6.0$\pm$0.18\,$\times$\,10$^{-15}$ & 0.82 \\
        \lbrack NII]$_{\lambda6548}$ & 7.3$\pm$0.15\,$\times$\,10$^{-15}$ & 1.01 \\
        H$\alpha_{\lambda6562.8}$ & 7.3$\pm$0.25\,$\times$\,10$^{-15}$ & 1.00 \\
        \lbrack NII]$_{\lambda6583}$ & 22.0$\pm$0.4\,$\times$\,10$^{-15}$ & 3.04 \\
        \lbrack SII]$_{\lambda6716}$ & 6.4$\pm$0.17\,$\times$\,10$^{-15}$ & 0.88 \\
        \lbrack SII]$_{\lambda6731}$ & 7.1$\pm$0.19\,$\times$\,10$^{-15}$ & 0.98 \\
        \hline
        \end{tabular}
\caption[Total fluxes from the extended emission]{Total fluxes of the measured emission lines from the extended emission in the core of the Centaurus cluster corrected for galactic and local extinction.  The \lbrack NII\rbrack$_{\lambda5755}$ flux is a 3$\sigma$ upper limit as denoted by the preceding $<$.}
\label{tab:ext}
\end{center}
\end{table}

Overall most of the measured lines fluxes from the extended emission are around an order of magnitude fainter than they are in the filaments. The exception here is the [OIII]$_{\lambda4959\&5007}$ emission line fluxes which are only a factor $\sim$3 fainter in the extended emission than the filaments.  Of particular note is that these lines are now significantly stronger relative to the other emission lines, being $\sim$ 5 times brighter relative to H$\alpha$ and $\sim$ 4 times brighter relative to [OI]$_{\lambda6300}$. Some of this difference may be a result of the increased uncertainty on the measured line fluxes due to the lower signal to noise, in particular the Balmer lines where the subtraction of continuum features compounds this.  However, we note that relative to H$\alpha$ the other line fluxes remain relatively consistent between the two regions, changing by no more than 30\%. The structure of the other forbidden lines in continuum subtracted narrowband images is similar to the [NII]$_{\lambda 6583}$ narrowband, showing some evidence of diffuse extended emission. However as these lines are significantly fainter it is not possible to map them to the full extent seen in [NII]$_{\lambda 6583}$.  

The relatively strong [OIII]$_{\lambda4959\&5007}$ emission suggests the physical conditions within these regions may be significantly different to that of the filaments.  In particular the high flux of [OIII]$_{\lambda4959\&5007}$ relative to [OI]$_{\lambda6300}$ largely rules out excitation through particle heating \citep{fer09}.  The standard line ratio diagnostic plots for the extended emission are shown in Figure \ref{fig:bpt} as the purple triangle.  As expected, and much like the filaments, the extended emission line ratios are very inconsistent with star formation excitation but due to the strong [OIII]$_{\lambda4959\&5007}$ are also inconsistent with particle heating.  The measured line ratios are best matched with the shock models of \citet{all08} (, which is consistent with the strong [OIII]$_{\lambda4959\&5007}$ emission found in these regions.  We do note that the measured [NII]/H$\alpha$ and [SII]/H$\alpha$ ratios are far higher than those predicted by the shock models, however at these densities the shock models are only defined for solar abundances while the abundances of N and S in the core of the centaurus cluster rise to approximatly 2\,Z$_{\odot}$ \citep{sf02,sf06}. This would increase these ratios so could account for this discrepancy.

\begin{figure}
\psfig{figure=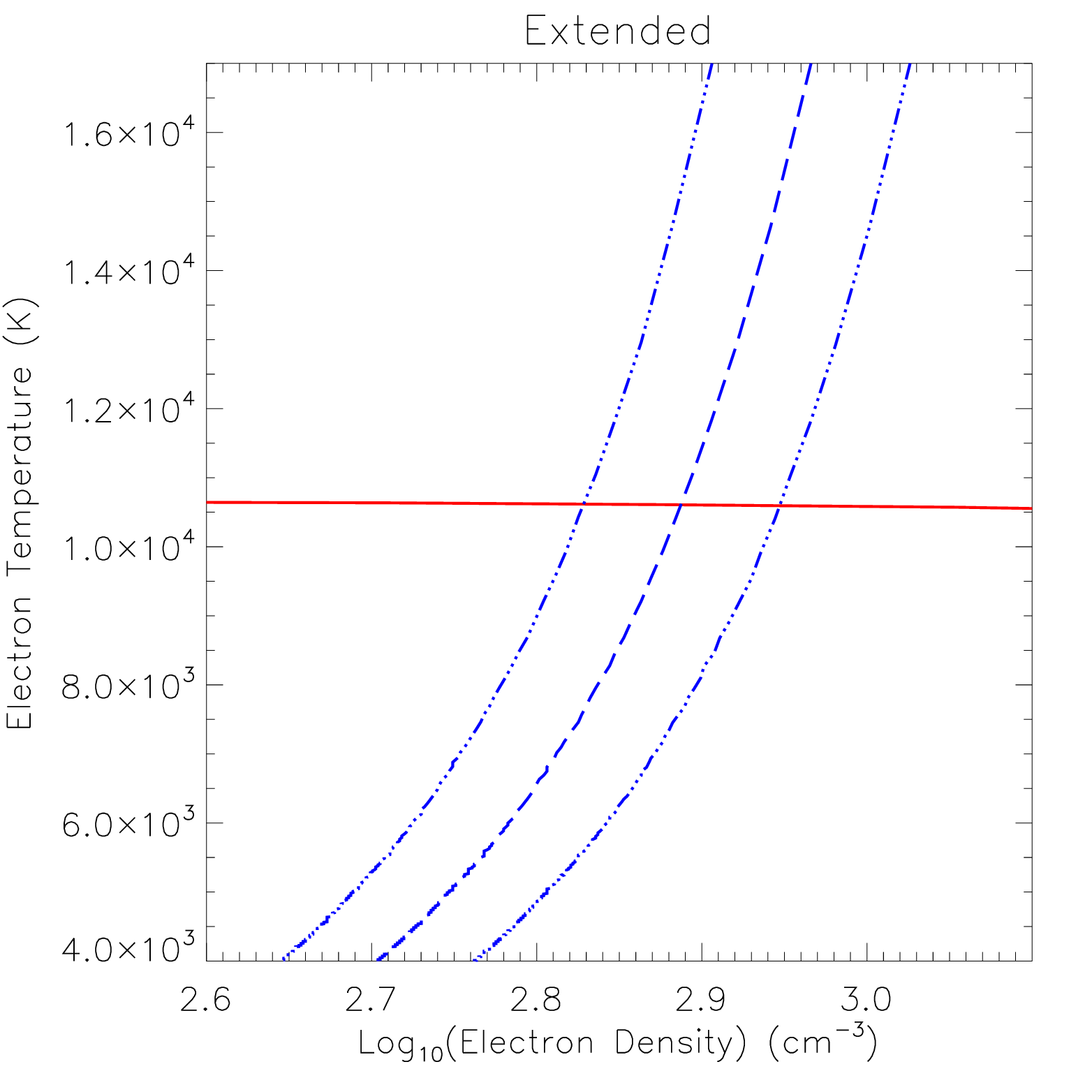,width=8.5cm}
\caption{Comparison plot of electron temperature and density within the extended emission as measured from the [NII]$_{\lambda5755}$ / ([NII]$_{\lambda6548}$ + [NII]$_{\lambda6583}$) (red solid line) and [SII]$_{\lambda6716}$ / [SII]$_{\lambda6731}$ (blue dashed line) emission line ratios.  The electron density is well constrained by the [SII]$_{\lambda6716}$ / [SII]$_{\lambda6731}$ ratio with the 1$\sigma$ error bounds shown by the blue triple dot dashed lines. While the electron temperature cannot be measured as the [NII]$_{\lambda5755}$ line is undetected, upper limits can be placed on the [NII]$_{\lambda5755}$ flux.  In this way the electron temperature can still be constrained to T$_e<$\,10555\,K.}
\label{fig:ext_temden}
\end{figure}

The [SII]$_{\lambda6716}$\,/\,[SII]$_{\lambda6731}$ line ratio measured from the extended emission is 0.89$\pm$0.034 making it significantly different to that of the filaments (1.16$\pm$0.021).  The [SII]$_{\lambda6716\&6731}$ lines are both detected at high significance allowing this ratio to be used to place strong constrains on the electron density (n$_e$) from these extended emission regions. However, the lower signal to noise means that the upper limit on the [NII]$_{\lambda5755}$ flux is significantly higher relative to the measured flux of the [NII]$_{\lambda6548\&6583}$ lines ($\sim$1\%).  As such the electron temperature (T$_e$) in this region is not as well constrained as in the filaments with a value of T$_e<$\,10555\,K.  The electron density within this region is then n$_e$=471--766\,cm$^{-3}$ (assuming a lower limit on temperature of 3500\,K) which is considerably higher than within the filaments (Figure \ref{fig:ext_temden}).  This puts an upper limit on the pressure of P$_e<$\,8.09$\times$10$^{6}$\,K\,cm$^{-3}$, significantly higher than within the filaments which is consistent with the presence of shocks within these regions.  Unfortunately as these values are measured from irregular regions between the bright filaments accurately matched estimates of the ICM pressure (which tend to be extracted from large radial bins) are not available.

\subsection{Emission from the filament free region to the north east}

\begin{figure}
\psfig{figure=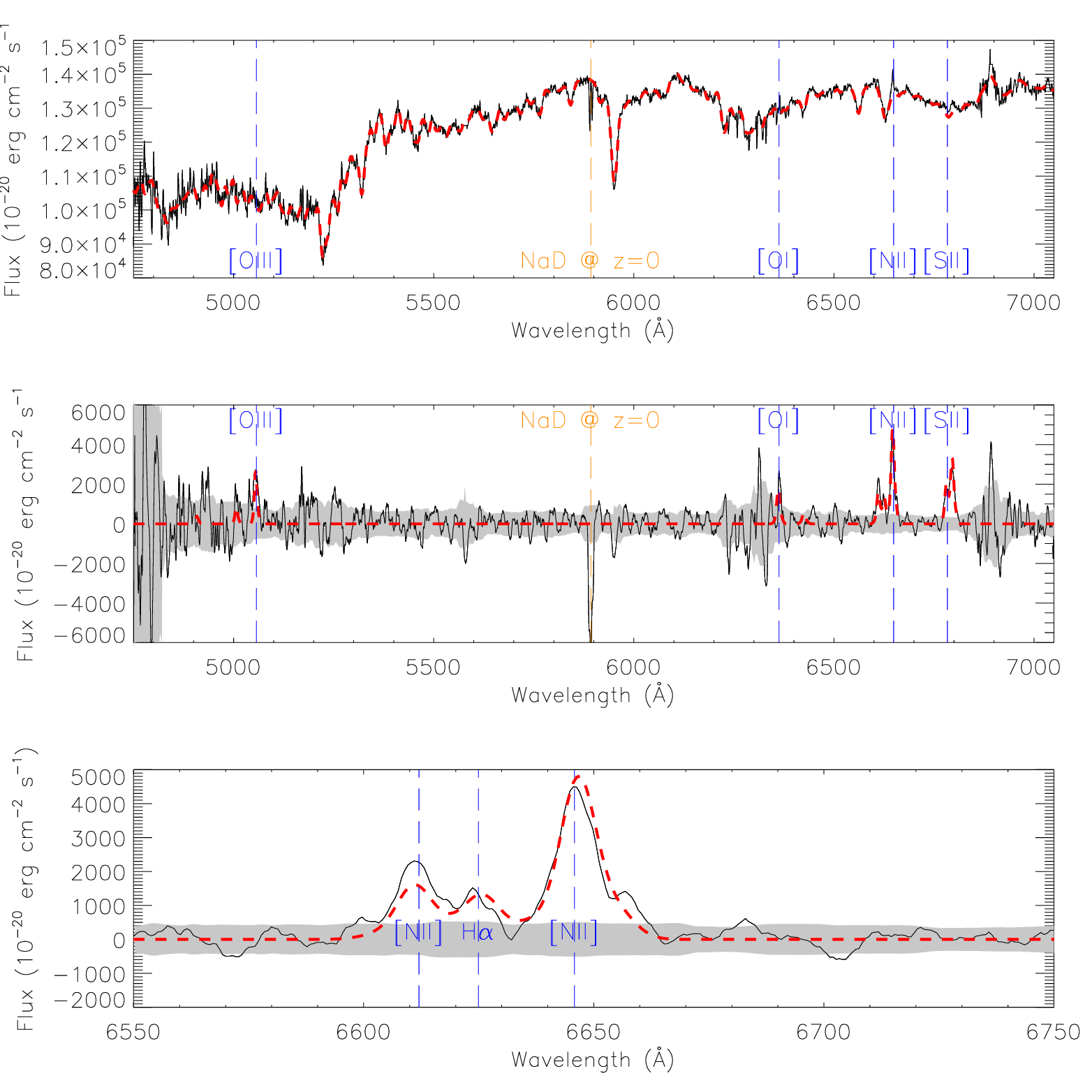,width=8.5cm}
\vspace{-0.75cm}
\caption[Spectrum and fit from the north east]{{\em Top:} Spectrum extracted from region A of the datacube to the north east of the BCG where no filaments are present.  {\em Centre:} Continuum subtracted spectrum extracted from region A.  {\em Bottom:} Continuum subtracted spectrum extracted from region A, showing only the region around [NII] and H$\alpha$.  The red dashed lines show the fits to the continuum (top) and emission line (centre and bottom) features of the spectrum.  The grey shaded region shows the 1$\sigma$ noise on the spectrum as a function of wavelength. [NII]$_{\lambda 6583}$ is visible in the spectrum prior to continuum subtraction, other lines typical of cluster cores are also detected.}
\label{fig:ne}
\end{figure}

As discussed in Section \ref{sec:res} the narrow band [NII]$_{\lambda 6583}$ image (Figure \ref{fig:nbs}) shows emission extended to the north of the BCG where no filaments have been detected in previous studies.
The structure of this emission closely matches that of a shell like structure in the x-ray emission seen by \citet{snd16} and corresponds to a high pressure region of the ICM (see their figure 13).
We designate this as region A and it occupies an area on the sky of $\sim$36\,arcsec$^2$.  Figure \ref{fig:ne} shows the extracted spectrum of Region A along with spectral fits to the stellar and nebula components.  Prior to the continuum subtraction only the [NII]$_{\lambda 6583}$ is visible in the spectrum but the emission lines detected in the other regions are revealed once the continuum is removed.  The measured fluxes of all lines detected in region A are listed in Table \ref{tab:rgA}.

\begin{table}
\begin{center}
\centerline{\sc Table. \ref{tab:rgA}.}
\centerline{\sc Total emission line fluxes from region A}
	\begin{tabular}{c c c}
    	\hline
		\noalign{\smallskip} 
        Emission & Total Flux & Flux/F$_{H\alpha}$ \\
        Line & (erg\,s$^{-1}$\,cm$^{-2}$) & \\
        \hline
        H$\beta_{\lambda4861}$ & 8.9$\pm$1.8\,$\times$\,10$^{-17}$ & 0.35 \\
        \lbrack OIII]$_{\lambda4959}$ & 26.5$\pm$2.0\,$\times$\,10$^{-17}$ & 1.03 \\
        \lbrack OIII]$_{\lambda5007}$ & 79.4$\pm$6.1\,$\times$\,10$^{-17}$ & 3.10 \\
        \lbrack NII]$_{\lambda5755}$ & $<$6.6\,$\times$\,10$^{-17}$ & 0.25 \\
        \lbrack OI]$_{\lambda6300}$ & 37.8$\pm$3.7\,$\times$\,10$^{-17}$ & 1.48 \\
        \lbrack NII]$_{\lambda6548}$ & 34.8$\pm$1.1\,$\times$\,10$^{-17}$ & 1.01 \\
        H$\alpha_{\lambda6562.8}$ & 25.6$\pm$5.2\,$\times$\,10$^{-17}$ & 1.00 \\
        \lbrack NII]$_{\lambda6583}$ & 104.4$\pm$3.3\,$\times$\,10$^{-17}$ & 4.08 \\
        \lbrack SII]$_{\lambda6716}$ & 43.9$\pm$3.3\,$\times$\,10$^{-17}$ & 1.72 \\
        \lbrack SII]$_{\lambda6731}$ & 70.8$\pm$3.3\,$\times$\,10$^{-17}$ & 2.77 \\
        \hline
        \end{tabular}
\caption[Total fluxes from region A]{Total fluxes of the measured emission lines from region A in the core of the Centaurus cluster corrected for galactic and local extinction.  The \lbrack NII\rbrack$_{\lambda5755}$ flux is a 3$\sigma$ upper limit as denoted by the preceding $<$.}
\label{tab:rgA}
\end{center}
\end{table}

The measured line fluxes from this region are all 2--3 orders of magnitude fainter than the total measured from the filaments, though we do note that region A covers a significantly smaller area.  While [NII]$_{\lambda6583}$ remains the brightest line the [OIII]$_{\lambda4959\&5007}$ lines are extremely bright relative to the other lines, with the [OIII]$_{\lambda5007}$ the second brightest having 76\% the flux of the [NII]$_{\lambda6583}$. The Balmer lines are very faint in this region, possibly due to the continuum fit not fully accounting for the absorption features, with an [NII]$_{\lambda6583}$/H$\alpha$ ratio of $>$4.  Despite this they remain consistent with each other, with H$\beta$/H$\alpha$ = 0.35, similar to that seen in the rest of the nebula.  The [NII]$_{\lambda 6583}$ flux detected from this region represents less than 0.5\% of the total from the nebula.  This results in an average surface brightness of 2.9$\pm$0.1$\times$10$^{-17}$\,erg\,cm$^{-2}$\,s$^{-1}$\,arcsec$^{-2}$ or 8.7$\pm$0.4$\times$10$^{-19}$\,erg\,cm$^{-2}$\,s$^{-1}$\,spaxel$^{-1}$ making it impossible to map the detailed structure of this extended emission with the data presented here.  

The strong [OIII]$_{\lambda4959\&5007}$ emission, with [OIII]$_{\lambda5007}$ more than three times brighter than the H$\alpha$ line and twice as bright as the [OI]$_{\lambda6300}$, suggests the physical conditions in this regions are more like that of the extended emission than the the filaments which is consistent with the faint and apparently diffuse structure of the region.  Line ratio diagnostics for region A are shown in Figure \ref{fig:bpt} as the cyan cross and immediately rule out excitation by star formation.  Much like the extended regions the line ratios are inconsistent with particle heating as the source of excitation within region A which is expected given the high [OIII]$_{\lambda5007}$/[OI]$_{\lambda6300}$ ratio.  The strong [OIII]$_{\lambda4959\&5007}$ suggests that shocks could be involved in exciting the gas, however the [NII]$_{\lambda6583}$/H$\alpha$ and [SII]$_{\lambda6716\&6731}$/H$\alpha$ ratios are too high to be explained by the shock models of \citet{all08}. To bring the measured values into agreement with the shock models would require either the gas abundances to be much higher than solar or a considerable underestimation of the measured Balmer line flux.

\begin{figure}
\psfig{figure=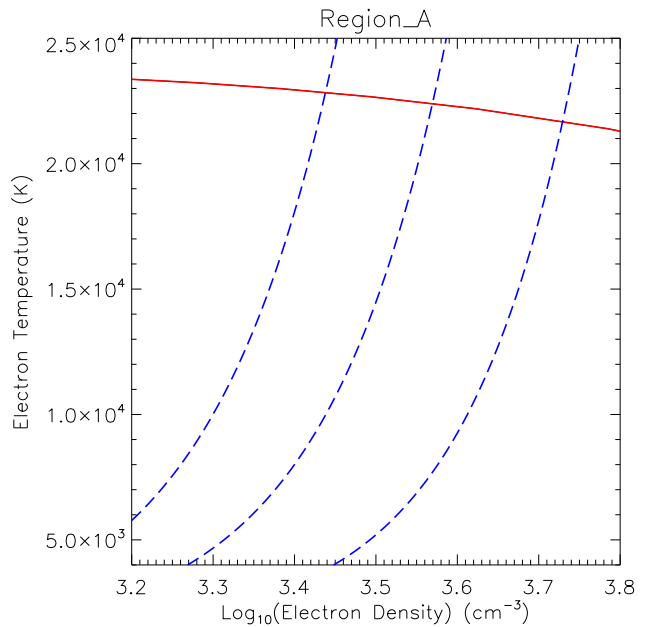,width=8.5cm}
\caption{Comparison plot of electron temperature and density within region A as measured from the [NII]$_{\lambda5755}$ / ([NII]$_{\lambda6548}$ + [NII]$_{\lambda6583}$) (red solid line) and [SII]$_{\lambda6716}$ / [SII]$_{\lambda6731}$ (blue dashed line) emission line ratios.  The electron density is well constrained by the [SII]$_{\lambda6716}$ / [SII]$_{\lambda6731}$ ratio with the 1$\sigma$ error bounds shown by the blue triple dot dashed lines. While the electron temperature cannot be measured as the [NII]$_{\lambda5755}$ line is undetected, upper limits can be placed on the [NII]$_{\lambda5755}$ flux.  In this way the electron temperature can still be constrained to T$_e<$\,22437\,K.}
\label{fig:rgA_temden}
\end{figure}

A key point to consider here however, is that region A occupies a high pressure region of the ICM \citep[see Figure 13 in][]{snd16} and matches the structure of a shell surrounding the radio source which \citet{snd16} suggest may be a shock generated by an outburst from the central AGN.  The close correlation between these two structures indicate the two are likely related, further suggesting shock excitation within region A. The [SII]$_{\lambda6716}$ / [SII]$_{\lambda6731}$ ratio in region A is significantly lower than other parts of the nebula at 0.62$\pm$0.054 indicating the emitting gas is significantly denser. Unfortunately despite remaining undetected the best upper limit on the [NII]$_{\lambda5755}$ flux for region A is very high at $\sim$6\% of the [NII]$_{\lambda6583}$ flux meaning the electron temperature in the region is very poorly constrained at T$_e<$\,22438\,K. Assuming a lower limit on temperature of 3500\,K this indicates an electron density of n$_e$=1729--3354\,cm$^{-3}$ (Figure \ref{fig:rgA_temden}) and a pressure of P$_e<$\,7.52$\times$10$^7$\,K\,cm$^{-3}$. Given the bright [OIII]$_{\lambda4959\&5007}$ emission, high density/pressure and matched shell like structure in the ICM, the production of line emission in region A through shock excitation is the explanation most consistent with the currently available data.

Despite this we note that the measured upper limit to the pressure (P$_e<$\,7.52$\times$10$^7$\,K\,cm$^{-3}$) is considerably higher than the pressure seen in the ICM which peaks at $\sim$0.1\,keV\,cm$^{-3}$ \citep{snd16} or $\sim$1.16$\times$10$^6$\,K\,cm$^{-3}$.  However, we note that the electron density, and thus the pressure, measured from the [SII]$_{\lambda6731}$ / [SII]$_{\lambda6716}$ ratio is dominated by the brightest (and densest) regions while the measured ICM pressure is the average over a large region.  This suggests that the line emitting gas in region A may be clumpy. Assuming that region A is related to the shell like structure seen in the ICM then its three dimensional structure can be fairly well constrained (the shell will expand roughly equally in all directions).  As such we can use the observed surface brightness of H$\alpha$ in region A to determine the average density using the emission measure method \citep[equations 3-35 \& 3-36,][]{spi78}. 

\begin{figure}
\vspace{0.5cm}
\psfig{figure=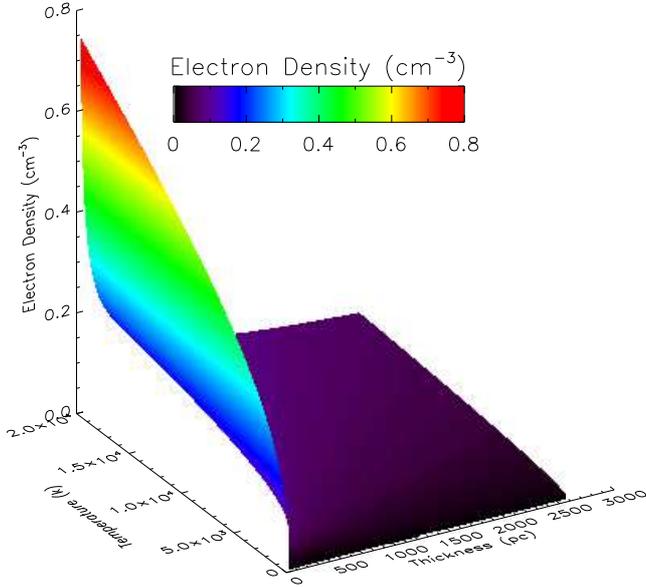,width=8.5cm}
\caption{Comparison plot of average electron density in region A as measured from the H$\alpha$ surface brightness and emission measure assuming a thickness of 16--2560\,pc.  The average electron densities are far lower than the value measured from the emission line ratios suggesting that either the thickness of the emitting medium is extremely low or the gas is clumpy with a low filling factor.}
\label{fig:em_n2}
\end{figure}

The measured intrinsic H$\alpha$ surface brightness of region A is 3.91$\pm$0.15$\times$10$^{-14}$\,W\,cm$^{-2}$\,sr$^{-1}$. Assuming the shell is spherical then the average distance through it is equal to average projected extent which is $\sim$2\,kpc.  The average density of the region (n$_{e,av}$) is then n$_{e,av}\sim$0.07\,cm$^{-3}$ at T$_e$=20000\,K (close to the upper limit of T$_e$<22437\,K) or n$_{e,av}\sim$0.035\,cm$^{-3}$ at the electron temperature of the filaments (T$_e$=5684\,K). Figure \ref{fig:em_n2} shows the measured average electron density for the parameter space T$_e$=100--20000\,K and thickness L=16--2560\,pc.  This does not exceed an electron density of n$_e$=0.8\,cm$^{-3}$, still orders of magnitude lower than the value measured from the emission lines.

This suggests that the emitting medium is either an extremely thin sheet in the plane of the sky or very clumpy.  We calculate that to match the electron density from the two measures would require the thickness of the emitting medium to be on the order of L=1$\times$10$^{-6}$\,pc.  Such a thin sheet of emitting gas extending over $\sim$2\,kpc (nine orders of magnitude more) in the plane of the sky seems extremely unlikely.  In the case of a clumpy medium the upper limit on the volume filling factor of the warm ionised gas in region A is $f_{v,ion}<$4$\times$10$^{-5}$ which is also extremely low. Given this and the apparent shock indicated by the X-ray shell it is likely that the line emitting structure has elements of both, forming a thin, flocculent shell of dense clumps along the shock front.

\subsection{Radial variation}

\begin{figure}
\psfig{figure=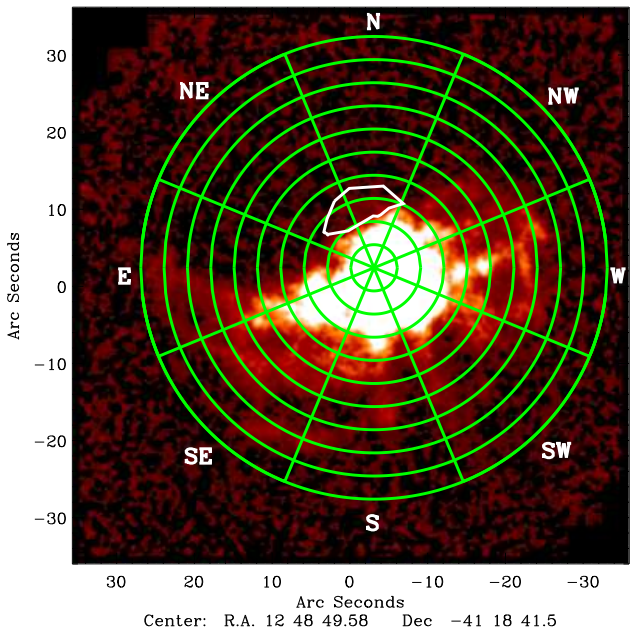,width=8.5cm}
\vspace{0.2cm}
\caption[\lbrack NII\rbrack narrowband image showing the radial bins from which spectra were extracted]{The [NII]$_{\lambda 6583}$ narrow band image showing the regions from which spectra were extracted and fitted to study the radial variations. The radial bins are numbered 1--10 from the centre out as shown in Figure \ref{fig:spcfits}. Region A is outlined in white for reference.}
\label{fig:rd}
\end{figure}

While the emission from this extended nebula is not sufficiently bright to resolve directly we can study its large scale properties through radial variations.  Spectra were extracted from radial bins (3" thick by 45$^\circ$ wide) along the cardinal and ordinal directions out to a distance of 30" ($\sim$6\,kpc) from the centre of the BCG (sufficient to cover the extended nebula and filaments, see Figure \ref{fig:rd}). The spectra from each bin were fitted and the parameters of the continuum and nebula line emission extracted.

\begin{figure}
\vspace{-0.35cm}
\psfig{figure=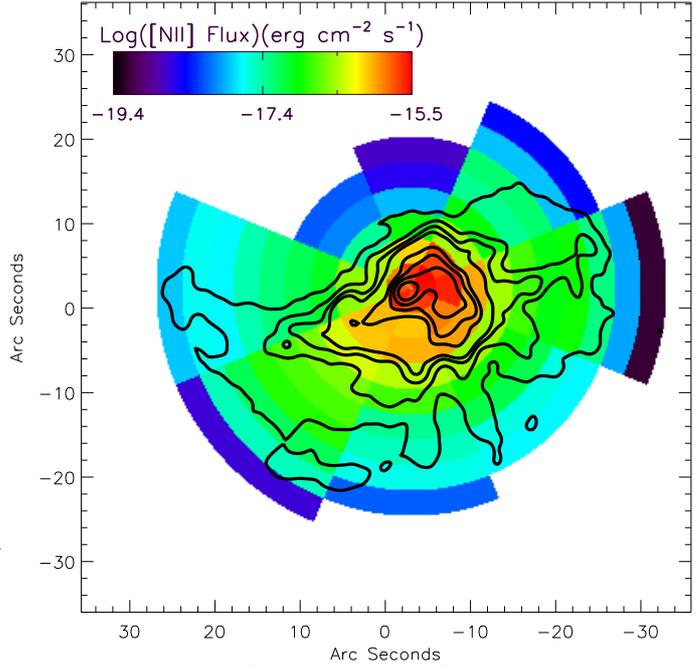,width=8.5cm}
\caption[Radial image of the detected \lbrack NII\rbrack flux]{Map of the detected [NII]$_{\lambda 6583}$ flux within the radial bins.   The [NII]$_{\lambda 6583}$ narrowband image showing the filamentary nebula is contoured in black (contours are at 2, 8, 14, 20, 50, 70, 100 and 150 $\times$10$^{-20}$\,erg\,cm$^{-2}$\,s$^{-1}$) for comparison.  [NII]$_{\lambda 6583}$ is detected out to $\sim$3.8 and 3.2\,kpc in the north and north east directions respectively, well beyond the extent of the structures seen in the narrow band images in those directions.}
\label{fig:rd_flux_img}
\end{figure}

\begin{figure}
\psfig{figure=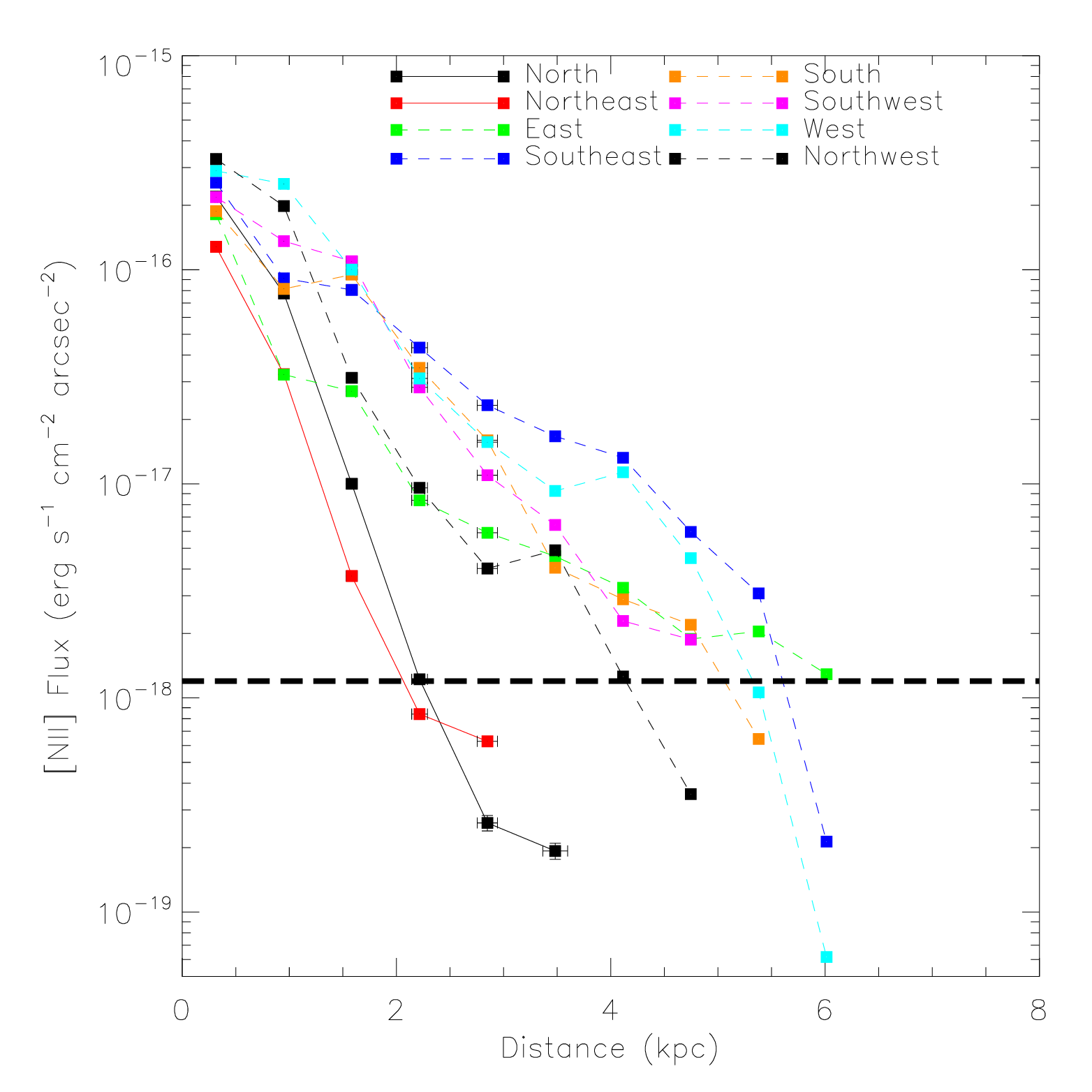,width=8.5cm}
\vspace{-0.75cm}
\caption[Radial profiles of the \lbrack NII\rbrack flux]{The radial variation of the [NII]$_{\lambda 6583}$ flux along cardinal and ordinal directions. The directions which contain filaments throughout are shown by the dashed lines while those with no filaments at large distances are shown by solid lines.  [NII]$_{\lambda 6583}$ is detected out to $\sim$3.8 and 3.2\,kpc in the north and north east directions respectively, well beyond the extent of the filamentary nebula in those directions.  Error bars are shown but in most cases are smaller than than the data point. We indicate the rms noise from the [NII]$_{\lambda 6583}$ narrowband image as the horizontal dashed line, points below this would not be detected in the image (assuming a uniform surface brightness within the radial bins).}
\label{fig:rd_flux}
\end{figure}

\begin{figure*}
\vspace{-0.35cm}
\psfig{figure=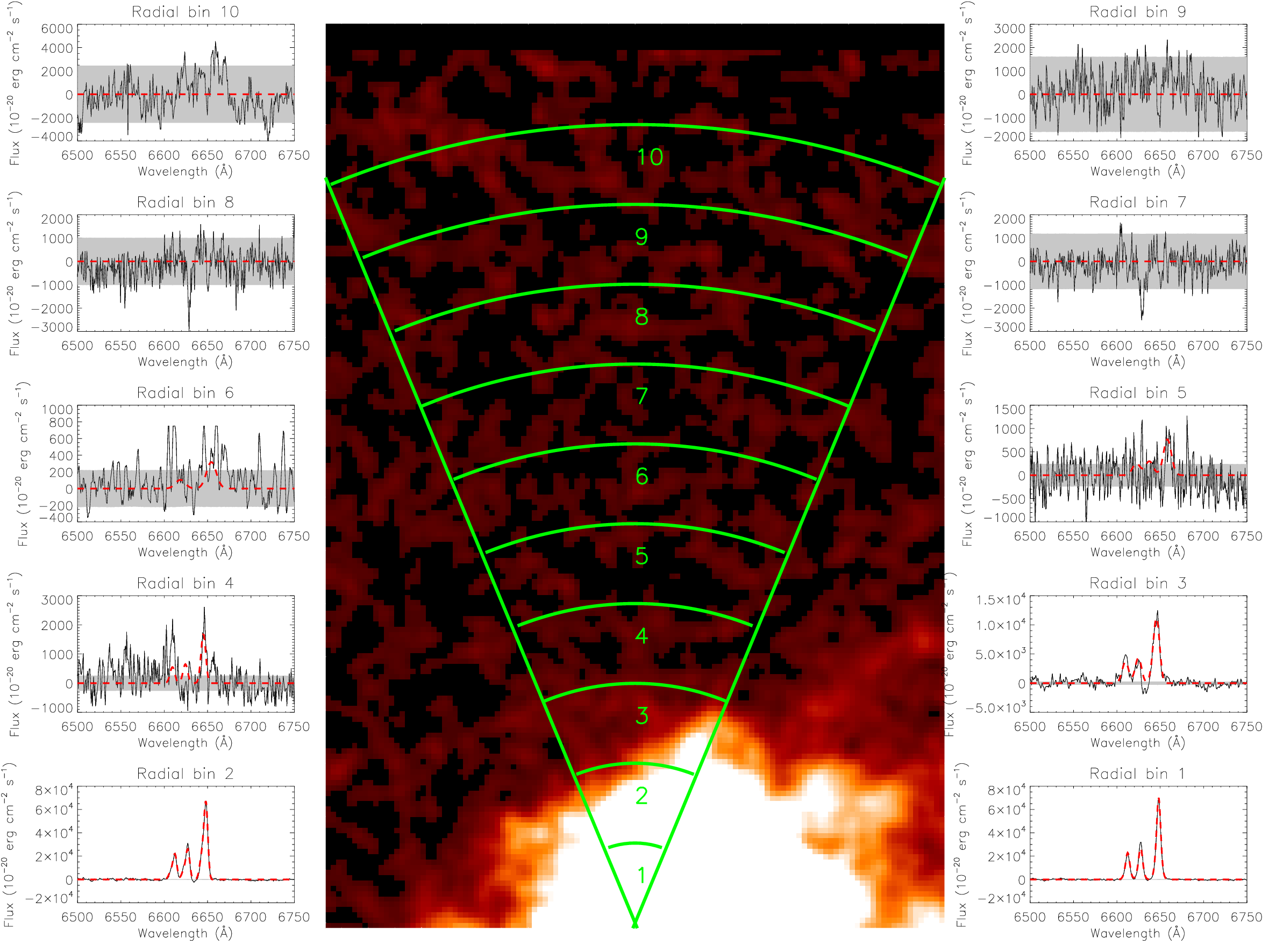,width=18cm}
\vspace{-0.3cm}
\caption[Radial detections of of \lbrack NII\rbrack to the north]{Plot showing the detections of [NII] \& H$\alpha$ from radial bins to the north of the BCG.  The central image shows the [NII]$_{\lambda 6583}$ narrow band image with the radial bins shown and numbered.  To either side (alternating right to left) are the continuum subtracted spectra (black solid lines) and emission line fits (red dashed lines) from each bin.  For reference the 1$\sigma$ noise at each spectral element is indicated by the grey shaded area.  Detections of [NII] are significant out to the 6th radial bin.}
\label{fig:spcfits}
\end{figure*}

In Figure \ref{fig:rd_flux_img} we present the [NII]$_{\lambda 6583}$ flux map produced from fits to these radial bins (radial flux profiles can be seen in Figure \ref{fig:rd_flux}). Along most directions [NII]$_{\lambda 6583}$ emission is only detected (at S/N\,>\,3) in radial bins which show some filamentary structure in the [NII]$_{\lambda 6583}$ narrow band image. However, we note that to the north and north east [NII]$_{\lambda 6583}$ is detected out to a distance of $\sim$3.8 and 3.2\,kpc (the 6th and 5th radial bins) respectively, extending beyond even region A (which mostly occupies the 3rd and 4th radial bins).  Figure \ref{fig:spcfits} shows the fits to the continuum subtracted spectra for the radial bins along the northern direction.  The [NII]$_{\lambda 6583}$ is clearly visible peaking above the noise level out to the 6th radial bin indicating the significance of these detections.

To the north and north east, [NII]$_{\lambda 6583}$ emission is detected in regions which do not contain bright filamentary structures (which only extend as far as $\sim$1.9 and 1.3\,kpc or the 3rd and 2nd radial bins) or any of the fainter structures seen in the [NII]$_{\lambda 6583}$ narrowband image (Figure \ref{fig:nbs} and discussed in the previous sections).  This emission indicates the presence of a nebula component which extends beyond both the filaments and the other extended structures already discussed. A similar structure is potentially present throughout the nebula but is not clearly detected in other regions due to the dominance of the emission from the filaments.  Indeed we note that along the north and north east directions the [NII]$_{\lambda 6583}$ flux is consistent with that detected along the directions containing filaments out to the second radial bin ($\sim$1.3\,kpc). Beyond this the flux drops sharply and is more than an order of magnitude fainter than the filaments by the 4th radial bin ($\sim$2.5\,kpc) and is consistent with the flux from the filaments at their most extended regions (8th-9th radial bins, $\sim$5-5.7\,kpc).  This suggests the most extended emission to the north and north-east may represent a more extended line emitting halo which the filaments are extending into.  


\begin{figure*}
\psfig{figure=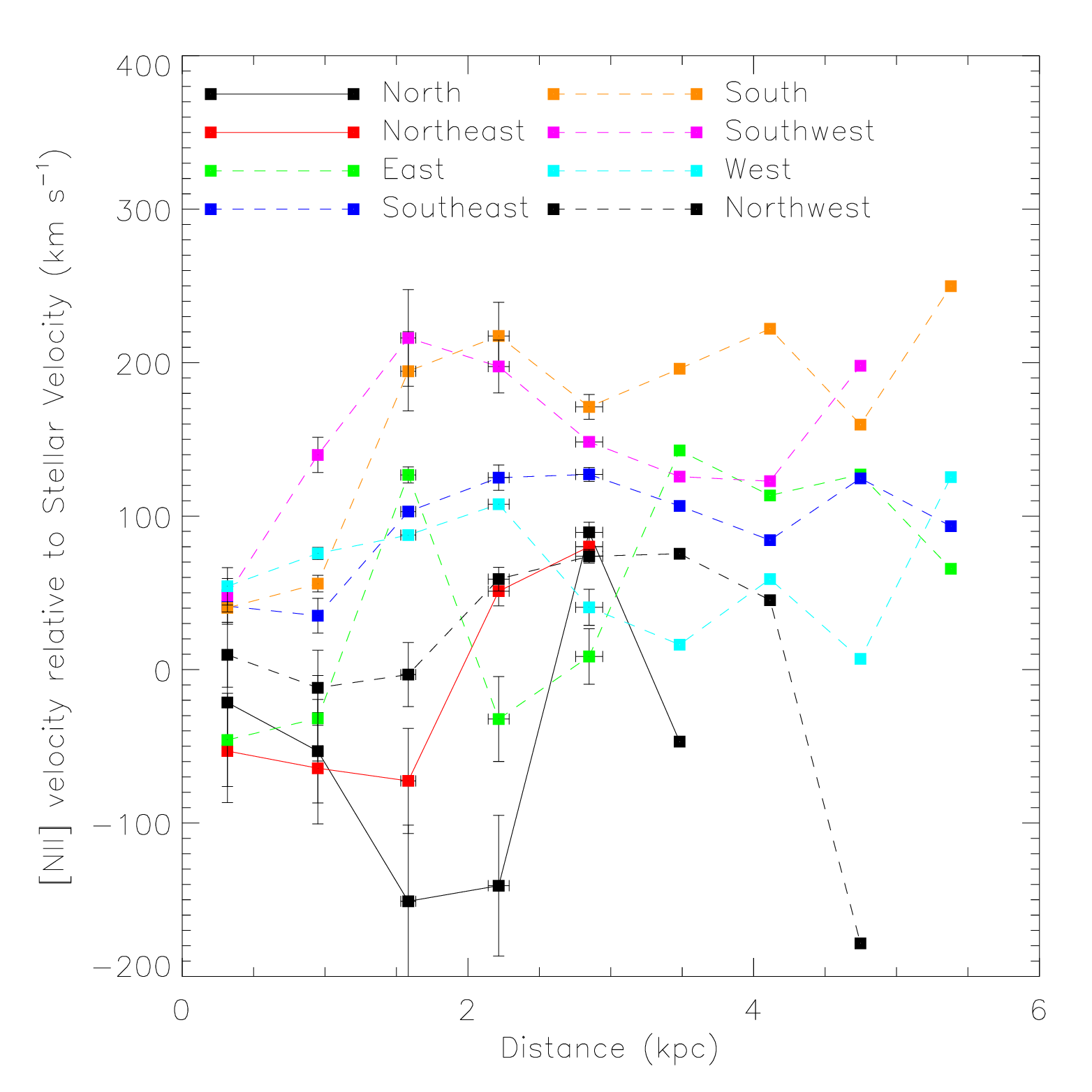,width=8.5cm}
\psfig{figure=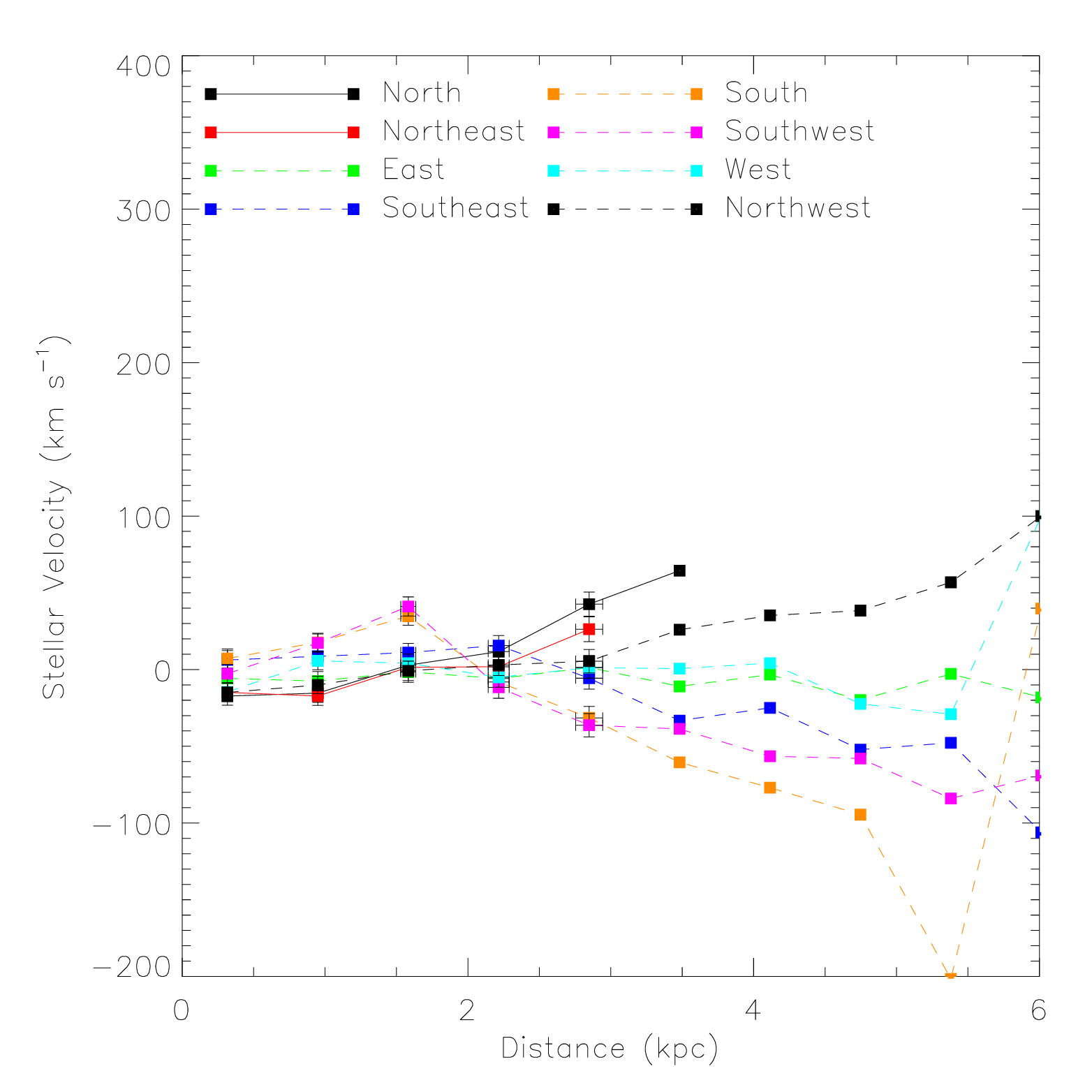,width=8.5cm}
\caption[Radial profiles of the \lbrack NII\rbrack velocity relative to the measured stellar velocity]{{\em Left:} The radial variation of the [NII]$_{\lambda 6583}$ velocity relative to the local stellar velocity along cardinal and ordinal directions. We note the large variation and offsets suggesting that the line emitting gas does not share the velocity of the BCG's extended stellar component.  We also note the strong contrast in velocities along opposing directions at $\sim$1.9\,kpc. In particular along the north-south and north east-south west directions hinting at a velocity structure within the gas. {\em Right:} Matching stellar velocities profiles (relative to the systemic velocity) derived from the full continuum fit shown with the same velocity range for comparison. The stellar velocity is relatively constant and consistent along all directions out to a distance of $\sim$2.21\,kpc from the centre of the BCG where they diverge slightly. This is not unexpected as BCGs typically show no ordered line of sight velocity structure in their stellar components and the relatively large bins will smooth out any small scale variations.}
\label{fig:rd_vdif}
\end{figure*}

\begin{figure}
\vspace{-0.35cm}
\psfig{figure=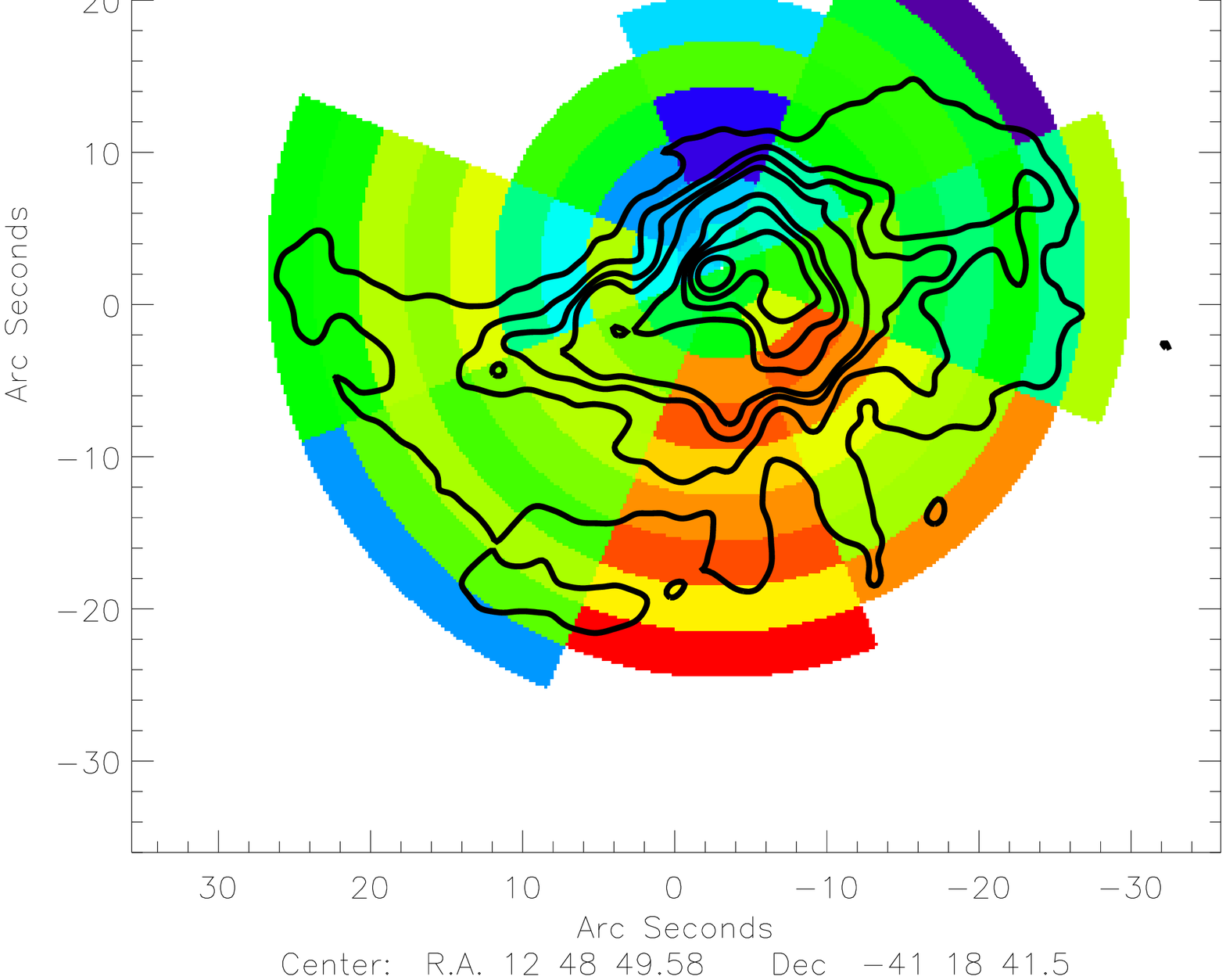,width=8.5cm}
\vspace{0.1cm}
\caption[Radial velocity map of the detected \lbrack NII\rbrack flux]{Map of the measured [NII]$_{\lambda 6583}$ velocity (relative to the local stellar velocity) within the radial bins.   The [NII]$_{\lambda 6583}$ narrowband image showing the filamentary nebula is contoured in black (contours are at 2, 8, 14, 20, 50, 70, 100 and 150 $\times$10$^{-20}$\,erg\,cm$^{-2}$\,s$^{-1}$) for comparison.  A strong velocity gradient can be seen along the north--south and north east--south west directions.}
\label{fig:rd_flux_vel}
\end{figure}

Profiles of the [NII]$_{\lambda 6583}$ velocity relative to the local stellar velocity (Figure \ref{fig:rd_vdif}) show considerable scatter but are typically significantly offset from a matching velocity indicating that both the filaments and halo are kinematically distinct from the BCGs extended stellar component. 
A clear velocity gradient can be seen in the resolved velocity map (Figure \ref{fig:rd_flux_vel}) running across the galaxy, from redshifted emission in the south and south--west to blueshifted in the north and north--east.  This was already seen in the velocity maps of \citet{can11}, though here the velocities are measured relative to that of the local stellar continuum (z=0.09869) making them $\sim$87\,km\,s$^{-1}$ higher.  However, to the north and north--east this velocity gradient reverses when it reaches the most extended regions detected. This can be seen as a shift back towards redshifted velocities of 377$\pm$13\,km\,s$^{-1}$ at $\sim$2.5\,kpc to the north and 328$\pm$50\,km\,s$^{-1}$ at $\sim$1.9\,kpc to the north--east and can also be seen in the velocity profiles (Figure \ref{fig:rd_vdif}).  At these distances the emission is from the halo and as such this velocity shift from filamentary to halo components may be indicating that the two are kinematically distinct. To confirm this it would be necessary to begin to resolve the diffuse emission so that the presence of a strong velocity shift across the transition could be confirmed.

\subsection{X-ray Comparison}

\begin{figure}
\psfig{figure=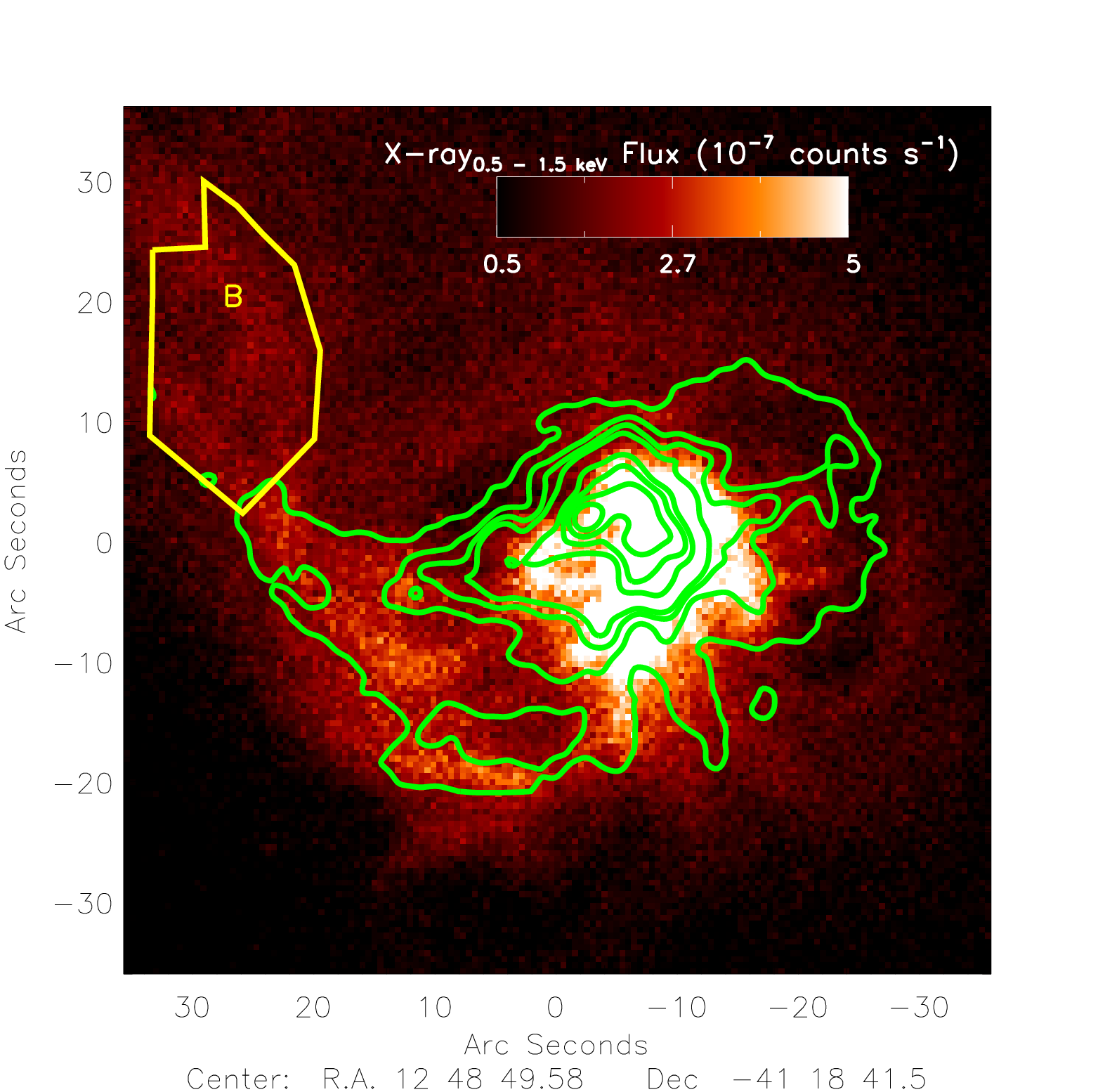,width=8.5cm,bbllx=96, bblly=401, bburx=576, bbury=883}
\vspace{0.2cm}
\caption[Soft X-ray image with \lbrack NII\rbrack contours]{The X-ray$_{0.5-1.5keV}$ image of the region of the Centaurus cluster covered by the MUSE observation with [NII]$_{\lambda 6583}$ contours (green) showing the strong similarities between the X-ray and optical line emitting structures.  Contours are at 2, 8, 14, 20, 50, 70, 100 and 150 $\times$10$^{-20}$\,erg\,cm$^{-2}$\,s$^{-1}$. We note however, region B (outlined in yellow) appears to be an extension of the X-ray filaments with no detected optical line emitting counterparts.}
\label{fig:xrn2}
\end{figure}

\begin{figure}
\psfig{figure=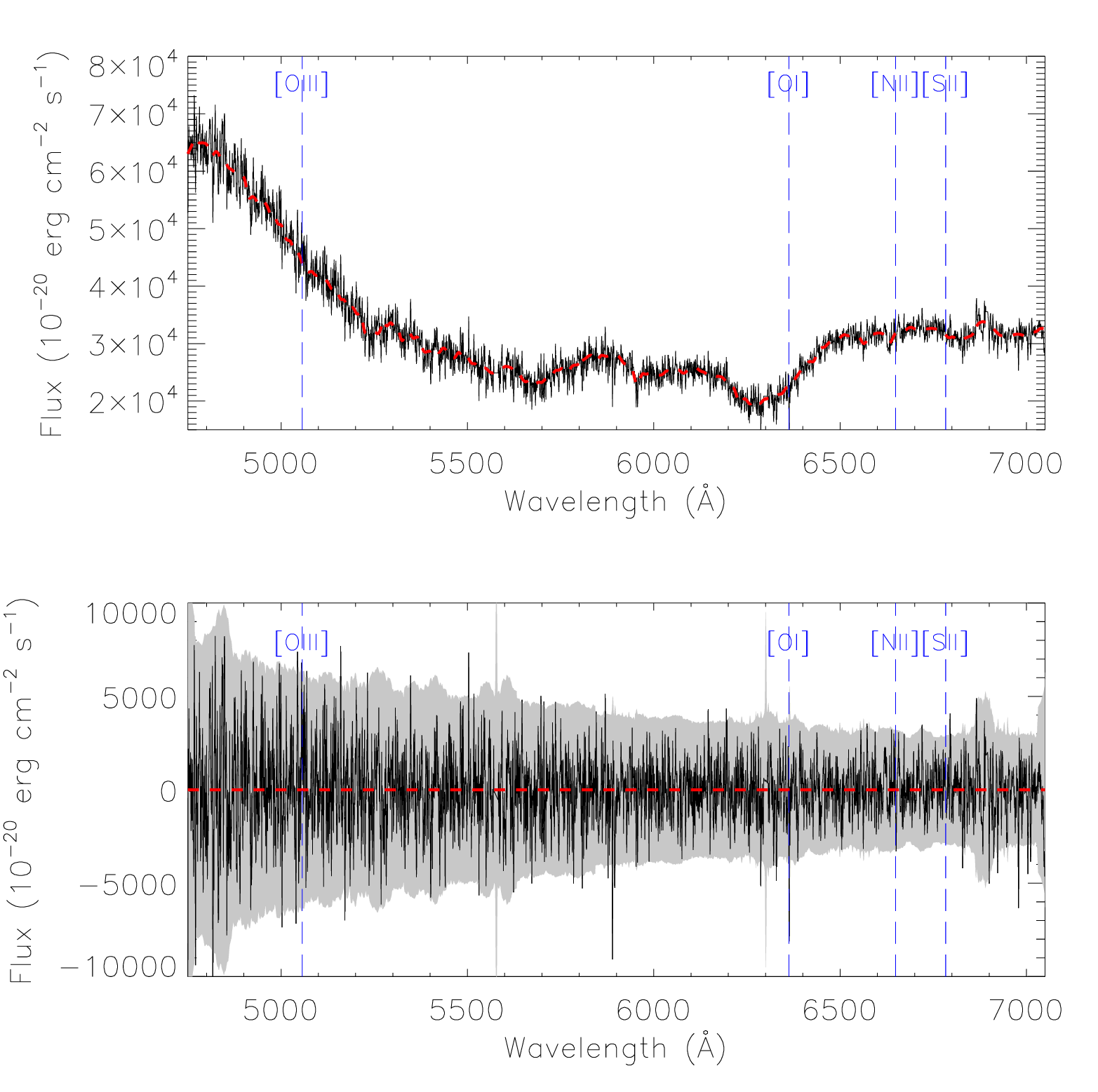,width=8.5cm}
\vspace{-0.75cm}
\caption[Spectrum and fit from the north east x-ray filaments]{{\em Top:} Spectrum extracted from the region of the MUSE datacube which corresponds to the north eastern X-ray filaments (Region B).  The dashed red line shows the continuum fit to the spectrum.  {\em Bottom:} Continuum subtracted spectrum extracted from the region of the datacube which corresponds to the north eastern X-ray filaments.  The red dashed line shows the fit to the emission line features of the spectrum. The spectrum lacks any emission line features typically found in cluster cores.}
\label{fig:nefil_spec}
\end{figure}

As is common in cluster cores the structure of the filamentary component of the optical nebula is qualitatively matched to the soft X-ray emission from the ICM (Figure \ref{fig:xrn2}).  Region B (outlined in Yellow) appears to break this trend showing bright X-ray filaments but no evidence of line emission even in its total spectrum (see Figure \ref{fig:nefil_spec}).  However, it must be noted that this region is at the edge of the MUSE cube so the signal to noise obtained will be lower due to having a lower total integration time (820--1640s) as a result of the dithering pattern employed during the observations.  The measured upper limit on the [NII] flux from this region is 34.5$\times$10$^{-17}$\,erg\,cm$^{-2}$\,s$^{-1}$ which is higher than the H$\alpha$ flux detected from region A where the X-ray surface brightness is similar.  As such it is possible that line emission is present in this region, but is below the detection limit of the current data.

What is unclear is if the X-ray\,--\,optical structural relation holds for the halo presented in this paper.  From Figure \ref{fig:xrn2} it is apparent that the X-ray emission does show some structure to the north and north east of the core (beyond region A), where the halo is most clearly detected. The brighter part of this strucutre extends roughly the same distance from the core as the detected [NII]$_{\lambda 6583}$ halo hinting that the halo may be following this strucutre. However this is far from conclusive, and would require the mapping of a significant fraction of the halo at comparable resolution to the ICM to confirm, something beyond the limits of the current data.

\begin{figure}
\psfig{figure=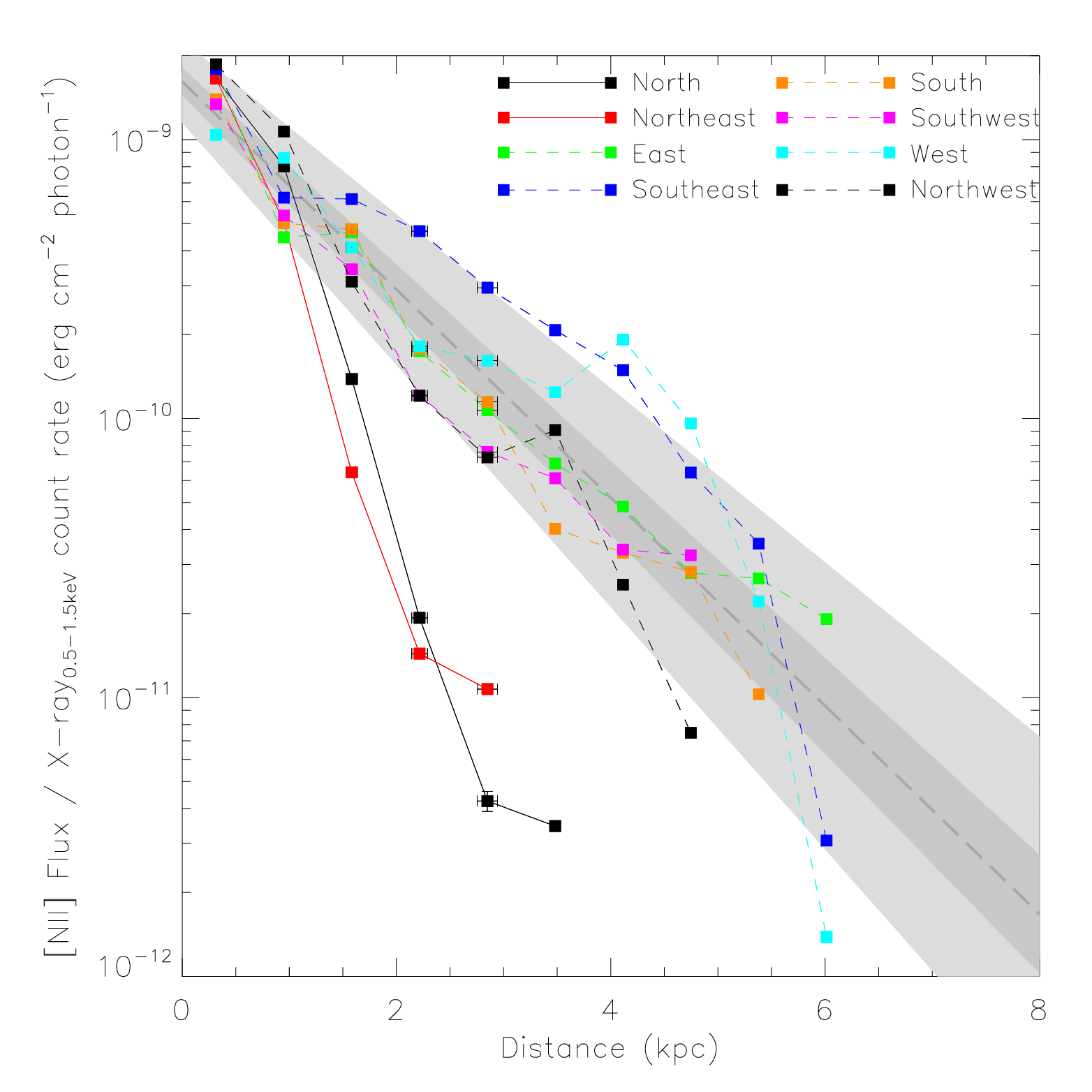,width=8.5cm}
\vspace{-0.75cm}
\caption[Radial profiles of the \lbrack NII\rbrack flux / X-ray count rate ]{The radial variation of the [NII]$_{\lambda 6583}$ Flux / X-ray$_{0.5-1.5keV}$ count rate along cardinal and ordinal directions. All profiles are highly consistent out to the second radial bin at $\sim$0.95\,kpc. Beyond this the profiles diverge with the regions containing filaments (dashed lines) having higher [NII]$_{\lambda 6583}$ / X-ray$_{0.5-1.5keV}$ ratios than the regions which contain no filaments (solid lines). The overall trend for the filaments is shown as the dashed grey line with the 1$\sigma$ (dark grey) and 3$\sigma$ (light grey) errors indicated.} 
\label{fig:rd_n2xy}
\end{figure}

Comparing the radial profiles of [NII]$_{\lambda 6583}$/X-ray$_{0.5-1.5keV}$ fluxes (Figure \ref{fig:rd_n2xy}) a general trend can be seen with the filamentary regions having higher ratios than the diffuse regions (region A and the Halo) at the same radius (similar to the [NII]$_{\lambda 6583}$ Flux profiles, Figure \ref{fig:rd_flux}). All profiles are tightly correlated out to the second radial bin ($\sim$0.95\,kpc). Beyond this the filaments and diffuse regions diverge with the filaments all following the general trend of the form $\frac{[NII]_{\lambda 6583}}{X-ray_{0.5-1.5keV}} = 10^{-0.47\pm0.03\times R -15.52\pm0.06}$ erg\,cm$^{-2}$\,photon$^{-1}$ (where R is the projected distance from the centre of the BCG in kpc) to within 3$\sigma$.  The diffuse regions fall more sharply and lie well below this trend suggesting either less optical line emitting gas is present or that it is excited through a different mechanism from the filaments.  This is consistent with our analysis of region A, which shows a very low gas filling factor and significantly different line ratios to the filaments. Unfortunately additional optical emission lines are to weak to be detected in the halo making studying the excitation and physical properties of the gas through standard line ratio diagnostics impossible with the current data.


\section{Discussion}

The optical line emission detected in extended and seemingly non filamentary regions of the Centaurus cluster can be classified into three different structures.

\begin{enumerate}[(a)]

\item {\bf Emission surrounding the filaments} - Line emission seen from the regions of the cluster core surrounding the bright and high contrast filaments.

\item {\bf Emission from the northern shell (region A)} - Line emission seen extending to the north of the BCG where there are no filaments and matching the position of an x-ray shell.

\item {\bf Emission from the extended halo} - Line emission seen only when high levels of binning are employed, seen to the north and north--east of the BCG beyond the northern shell.

\end{enumerate}

The emission from these structures appears to differ in many ways to the previously known emission from the filaments.  In particular the emission surrounding the filaments and the northern shell suggest the gas in these regions is much denser (by as much as an order of magnitude in the case of region A) than in the filaments. Similarly, our line ratio analysis suggests that the dominant excitation mechanism in these regions may be shocks while the values are much more consistent with particle heating in the filaments.  

Given the fact that the emission surrounding the filaments appears to follow the large scale filamentary structure of the nebula it seems unlikely that this emission is unrelated to the filaments.  Despite this its excitation state (as measured by the line ratios) is not consistent with the dominant heating mechanism in the filaments (particle heating).  However, the line ratios are more consistent with shock excitation, which is consistent with the higher electron density measured within these regions.  It seems likely then that this emission originates from shocked regions of gas outside the filaments, perhaps liberated from the filaments by motions of the ICM or deposited in situ outside of the magnetic fields which support the filaments.  Indeed the possibility of pressure imbalances within the multiphase ICM \citep{jaf01,jaf05} could present a possibility where the outer shells of the filaments expand to fill the regions between them. However, in such a case the density and pressure of this extended gas would be lower than that of the filaments which is in contrast to the values measured here. It must be noted though that the [SII]$_{\lambda6731}$ / [SII]$_{\lambda67316}$ line ratio is most sensitive to the densest clumps of the gas and that the mean density of this region is likely much lower than indicated given the low surface brightness.

Likewise, the striking similarity in the structure and extent of region A to the X-ray shell seen by \citet{snd16} suggests that the two structures are likely related.  Indeed \citet{snd16} suggest the shell may be related to a shock in the ICM and the high density of the line emitting gas in region A, as well as the line ratio diagnostics, are consistent with the presence of a shock in this region.  A similar situation can be found within the Perseus cluster where a weak shock to the north of the BCG which corresponds to a region devoid of filaments \citep{gra08}.  Interestingly in this case filaments are present either side of the shock, terminating at the shock front,suggesting that filaments may have once been continuous through this region but have since been disrupted by the shock.
The low volume filling factor in region A ($f_{v,ion}$\,<4$\times$10$^{-5}$) suggests that the gas is either extremely clumpy or forms a thin ($\sim$10$^{-6}$\,pc thick) bubble expanding with the shock front.  Given the extremely low values of both the volume filling factor and thickness of a continuous shell, the actual structure is potentially a combination, in the form of a thin clumpy regions expanding with the shock.

The nature of the extended halo is much less clear as it is so faint that only the brightest emission line ([NII]$_{\lambda 6583}$) can be significantly detected.  However, the fact that it can only be detected when the spectra are binned over very large extents suggests that the emission may be diffuse rather than concentrated in narrow regions as is seen in the filaments.  While the limits of the data make it difficult to draw strong conclusions on the nature of this halo emission several potential origins can still be explored:

\begin{enumerate}

\item {\bf Faint Filaments} could be present throughout the sampled region which are not sufficiently bright to be resolved. If the flux from these filaments was spread uniformly over $\sim$100 spaxels within each radial bin then this would make them comparable to the noise in our narrowband images and they would thus be undetected.  This represents $<$25\% of the bin size beyond the 2$^{nd}$ radial bin ($\sim$0.95\,kpc) so cannot be ruled out with the current data.

\item {\bf Gas from destroyed/disrupted filaments} that were once present in this region could now be spread more diffusely throughout the cluster core. Such filaments could have potentially been stripped by the ICM if a relative motion was present or could have been ``evaporated" through the excitation mechanisms which illuminate them.  Filaments are believed to be stabilised against such processes due to support by magnetic fields \citep{fab08} and this stabilisation would need to remain after the disruption of the filaments to prevent the cold gas being lost to the extremely hot ICM ($\sim$10$^7$\,K) and it is currently unclear how this could be accomplished.  

\item {\bf Mass loss from stars} in the BCGs extended envelope could be responsible for the gas producing this emission.  In this case however, we would expect the gas to share the local stellar velocity within each bin which does not appear to be the case (see Figure \ref{fig:rd_vdif}).

\item {\bf Gas cooling from the ICM} will recombine as it falls from 10$^6$\,K to 10$^3$\,K resulting in optical line emission, however the emission line flux detected from filaments is typically far in excess of what is expected to be produced in this way \citep{jst87}.  Within the radial bins from the halo regions the  [NII]$_{\lambda 6583}$ luminosity falls to $\sim$1.5$\times$10$^{37}$\,erg\,s$^{-1}$ at the most distant regions from the cluster.  Assuming the [NII]$_{\lambda 6583}$/H$\alpha$ ratio measured from the total nebula (2.33$\pm$0.06) and the standard H$\alpha$ to H$\beta$ ratio of 2.86 (case B recombination at T=10$^4$\,K \& n$_e$=10$^2$\,cm$^{-3}$ with no reddening) then using equation 1 from \citet{jst87} this line emission would represent a mass deposition rate of $\sim$0.1--0.2\,M$_\odot$\,yr$^{-1}$. While this seems low it represents a region of at most 3\,kpc$^2$ and if it were constant over the full region containing filaments would represent a total mass flow rate of 4.2--8.3\,M$_\odot$\,yr$^{-1}$ from the inner 6\,kpc region.  The lower end of this range is remarkably consistent with the upper limit of cooling determined from the FeXVII emission feature in the RGS spectra from a similar region of the Centaurus cluster core \citep[4\,M$_\odot$\,yr$^{-1}$][]{snd08}.  

\item {\bf Light scattered along the line of sight} by the ICM could be responsible for the extended diffuse emission \citep{fab89,hin99}.  Given the considerably lower flux in the diffuse regions this could be light which originated within the filaments. However, the electrons within the ICM will have thermal velocities of $\sim$10000\,km\,s$^{-1}$ (at 0.5\,keV) which would broaden the scattered light to comparable line widths \citep{khe14} which would be undetectable in the spectrum and are far in excess of those measured.  The possibility remains that line emission could be scattered by cooler gas or dust \citep[e.g.][]{dey96,zak05}, however dust should be quickly sputtered when exposed to the ICM unless shielded within dense cold gas clouds.  As such scattering could only be responsible if cool gas were already present and as such is likely only having a boosting effect on the emission from these regions.  To determine the fraction of emission that comes from scattered light would require polarization observations of these very faint regions which is beyond the capabilities of current instrumentation. 

\end{enumerate}

Since scattering would require cold gas to be present within these regions already option (v) cannot explain the observed features alone. The fact that the gas velocity does not match that of the stellar population makes it difficult to reconcile the line emission with gas produced by mass loss from stars in the BCGs extended envelope making origin (iii) extremely unlikely.  The cooling rates from the ICM do appear sufficient to account for the emission from the halo suggesting option (iv) could be at work here. However, we note that the halo emission is detected from only a small region of the cluster core while emission from cooling gas would be expected to be spread more uniformly.  This could potentially be a result of the bright filaments obscuring the halo emission in other parts of the cluster core, but it is not possible to resolve the individual components with the current data.

The radio structure from the AGN at the core of the Centaurus cluster shows a jet to the north splitting to form lobes which extend southwards forming a distinctive ``U" shape \citep[e.g.][]{ode94,tay06,mit11}.  Such a structure suggests the possibility that the BCG is moving north at a significant velocity, leaving the radio plumes trailing behind it.  This is consistent with the presence of a higher pressure region of the ICM to the north of the BCG. \citet{tay06} do suggest that the jets appear to be being produced close to the line of sight (<\,70$^\circ$ inferred from Doppler boosting) but also note that on larger scales the orientation of the jet is strongly modified so this is not inconsistent with the ``U" shape being produced by a moving BCG. Such a motion could produce a wind as the BCG moves through the ICM which could have destroyed any filaments to the north east, disrupting them, and spreading their gas into the ICM.  

The halo emission's velocity offset from the BCG (of the order 100\,km\,s$^{-1}$) and the extent over which the emission is detected (of the order 1\,kpc) would imply that even if this emission is related to disrupted filaments cold gas clouds must remain within this region for $\sim$10\,Myrs ($\frac{1\,kpc}{100\,km\,s^{-1}} = \frac{3.086\times10^{21}\,cm}{10^7\,cm\,s^{-1}} = 3.086\times10^{14}s \sim 10\,Myr$) in order to produce the emission detected. This suggests that the cold gas from any disrupted filaments could not be short lived and must therefore retain some protection from the hot gas in the ICM \citep[likely threaded by magnetic fields][]{fab08}. While it is not clear how this could be achieved this is not sufficient to rule out origin (ii).

Of the four origins proposed only the third can be ruled out from the current data. Options (i), (ii) and (iv) all remain possibilities, although none are without their challenges, while option (v) is potentially boosting the observed flux. Unfortunately distinguishing between these options would require mapping the global structure, resolving the small scale structure and measuring the flux of multiple emission lines of this halo emission which is beyond the limits of the current data.
%

\section{Conclusions}

We present the detection of previously unknown extended structures seen in [NII]$_{\lambda 6583}$ emission surrounding the BCG in the Centaurus cluster.  Notably these structures are much fainter than the filaments and appear to be more extended as well.  The first structure appears to be related to the filaments, surrounding them and filling the gaps between them indicating that the gaps between filaments are not devoid of cool gas.  A second structure forms a shell to the north of the BCG and is spatially coincident with an apparent shock structure seen in the X-ray emission.  This fact and the high density and line ratios suggest that this structure is likely part of a shocked shell of gas surrounding the radio source.

The final structure appears in the form of halo emission which is orders of magnitude fainter than that from the previously known filaments. It is offset from the nearby filaments by 100--300\,km\,s$^{-1}$ in the region where it is most strongly detected.  Noteably this offset is in the opposite direction to the larger scale velocity gradient (which produces similar sized offsets elsewhere in the nebula) suggesting it may be kinematically distinct. While it is too faint to extract measurable line ratios the measured [NII]$_{\lambda 6583}$ flux is consistent with that from filamentary regions with comparable X-ray flux, which does not suggest a drastically different excitation mechanism.  

We conclude that this structure is most consistent with a filamentary origin, either representing extremely faint filaments or emission from filaments which have been disrupted (and their gas dispersed more diffusely) by their interactions with the cluster.  However, we cannot rule out the possibility that this emission is related to gas cooling directly from the ICM and the fact that the required cooling rate to produce this emission is consistent with the measured cooling rate of the ICM is strongly suggestive of this possibility.  In order to distinguish between these options more sensitive (and thus deeper) observations are required to allow us to map this extended structure and measure line ratios to more accurately constrain the gas excitation. Such observations should also cover a more extended region of the cluster core to allow a more extensive search for halo emission and additional soft x-ray structures within the ICM (such as the filaments to the far north east) to be investigated.

The optical nebula within cluster cores are typically assumed to be filamentary in nature as their structure is almost always seen to be filamentary.  This may simply be due to the fact that the filaments make up the vast majority of the line flux and this discovery of a line emitting halo in the Centaurus cluster presents the first evidence that a more diffuse underlying structure may also be present.  However, with the data presented it is not possible to rule out very faint filaments as a potential source of this newly detected emission, so it remains possible that deeper observations may yet identify filamentary structures within the halo.
Despite the limitations, this discovery opens up the possibility of there being much more extended halos of optical line emission, and thus the associated cold gas, in cluster cores than previously known. 

\section*{Acknowledgements}

SLH \& ACF acknowledge support from the European Research Council for Advanced Grant Program num 340442--FEEDBACK.

PS, FC, VO, FP \& RB acknowledge support from the ANR grant LYRICS (ANR-16-CE31-0011).

Based on observations made with ESO Telescopes at the La Silla or Paranal Observatories under programme ID 094.A-0859.





\bibliographystyle{mnras}
\bibliography{bib}




\bsp	
\label{lastpage}
\end{document}